\documentclass[journal]{IEEEtran}
\usepackage{cite}
\usepackage{amssymb,amsmath}
\usepackage[dvips]{graphicx}
\usepackage[usenames]{color}
\usepackage{amsfonts}
\usepackage{fancyhdr}
\usepackage{subfig}
\usepackage{latexsym}
\usepackage{amssymb,amsthm,amsmath}
\def\be{\begin{equation}}

\def\bi{\begin{itemize}}
\def\ee{\end{equation}}
\def\ei{\end{itemize}}

\begin{document}
%
\title{Censored Truncated Sequential Spectrum Sensing for Cognitive Radio Networks}
%
%
%

\author{Sina~Maleki~
        ~Geert~Leus~
        \thanks{S. Maleki and G. Leus are with the Faculty of Electrical Engineering, Mathematics
and Computer Science, Delft University of Technology, 2628 CD Delft,
The Netherlands (e-mail: s.maleki@tudelft.nl; g.j.t.leus@tudelft.nl). {Part of this paper has been presented at the 17th International Conference on Digital Signal Processing, DSP 2011, July 2011, Corfu, Greece.} This work is supported in part by the NWO-STW under the VICI program (project 10382). Manuscript received date: Jan 5, 2012. Manuscript revised dates: May 16, 1012 and Jul 19, 2012}}%
%

%


\maketitle

\begin{abstract}
Reliable spectrum sensing is a key functionality of a cognitive radio network. Cooperative spectrum sensing improves the detection reliability of a cognitive radio system but also increases the system energy consumption which is a critical factor particularly for low-power wireless technologies. A censored truncated sequential spectrum sensing technique is considered as an energy-saving approach. To design the underlying sensing parameters, the maximum {average energy consumption per sensor} is minimized subject to a lower bounded global probability of detection and an upper bounded false alarm rate. This way both the interference to the primary user due to miss detection and the network throughput as a result of a low false alarm rate are controlled. {To solve this problem, it is assumed that the cognitive radios and fusion center are aware of their location and mutual channel properties.} We compare the performance of the proposed scheme with a fixed sample size censoring scheme under different scenarios and show that for low-power cognitive radios, censored truncated sequential sensing outperforms censoring. It is shown that as the sensing energy per sample of the cognitive radios increases, the energy efficiency of the censored truncated sequential approach grows significantly.
  
\end{abstract}


%
\IEEEpeerreviewmaketitle

\section{Introduction}

Dynamic spectrum access based on cognitive radios has been proposed in order to opportunistically use underutilized spectrum portions of the licensed electromagnetic spectrum \cite{ZS}. 
Cognitive radios opportunistically share the spectrum while avoiding any harmful interference to the primary licensed users. They employ spectrum sensing to detect the empty portions of the radio spectrum, also known as spectrum holes. Upon detection of such a spectrum hole, cognitive radios dynamically share this hole. However, as soon as a primary user appears in the corresponding band, the cognitive radios have to vacate the band. As such, reliable spectrum sensing becomes a key functionality of a cognitive radio network.

The hidden terminal problem and fading effects have been shown to limit the reliability of spectrum sensing. Distributed cooperative detection has therefore been proposed to improve the detection performance of a cognitive radio network \cite{DCK}, \cite{MSB}. Due to its simplicity and small delay, a parallel detection configuration \cite{V}, is considered in this paper where each secondary radio continuously senses the spectrum in periodic sensing slots. A local decision is then made at the radios and sent to the fusion center (FC), which makes a global decision about the presence (or absence) of the primary user and feeds it back to the cognitive radios. Several fusion schemes have been proposed in the literature which can be categorized under soft and hard fusion strategies \cite{V}, \cite{Kay}.
Hard schemes are more energy efficient than soft schemes,  and thus a hard fusion scheme is adopted in this paper. {More specifically, two popular choices are employed due to their simple implementation: the OR and the AND rule. The OR rule dictates the primary user presence to be announced by the FC when at least one cognitive radio reports the presence of a primary user to the FC. On the other hand, the AND rule asks the FC to vote for the absence of the primary user if at least one cognitive radio announces the absence of the primary user.} In this paper, energy detection is employed for channel sensing which is a common approach to detect unknown signals \cite{Kay}, \cite{CMB}, and which leads to a comparable detection performance for hard and soft fusion schemes \cite{MSB}.

Energy consumption is another critical issue. The maximum energy consumption of a low-power radio is limited by its battery. As a result, energy efficient spectrum sensing limiting the maximum energy consumption of a cognitive radio in a cooperative sensing framework is the focus of this paper.

\subsection{Contributions}
The spectrum sensing module consumes energy in both the sensing and transmission stages. To achieve an energy-efficient spectrum sensing scheme the following contributions are presented in this paper.
\begin{itemize}
\item A combination of censoring and truncated sequential sensing is proposed to save energy. The sensors sequentially sense the spectrum before reaching a truncation point, $N$, where they are forced to stop sensing. If the accumulated energy of the collected sample observations is in a certain region (above an upper threshold, $a$, or below a lower threshold, $b$) before the truncation point, a decision is sent to the FC. Else, a censoring policy is used by the sensor, and no bits will be sent. This way, a large amount of energy is saved for both sensing and transmission. {In our paper, it is assumed that the cognitive radios and fusion center are aware of their location and mutual channel properties.}

\item Our goal is to minimize the maximum {average energy consumption per sensor} subject to a specific detection performance constraint which is defined by a lower bound on the global probability of detection and an upper bound on the global probability of false alarm. In terms of cognitive radio system design, the probability of detection limits the harmful interference to the primary user and the false alarm rate controls the loss in spectrum utilization. The ideal case yields no interference and full spectrum utilization, but it is practically impossible to reach this point. Hence, current standards determine a bound on the detection performance to achieve an acceptable interference and utilization level \cite{SCSC}. To the best of our knowledge such a min-max optimization problem considering the {average energy consumption per sensor} has not yet been considered in literature.  

\item Analytical expressions for the underlying parameters are derived and it is shown that the problem can be solved by a two-dimensional search {for both the OR and AND rule.}

\item To reduce the computational complexity for the OR rule, a single-threshold truncated sequential test is proposed where each cognitive radio sends a decision to the FC upon the detection of the primary user. 

\item To make a fair comparison of the proposed technique with current energy efficient approaches, a fixed sample size censoring scheme is considered as a benchmark (it is simply called the censoring scheme throughout the rest of the paper) where each sensor employs a censoring policy after collecting a fixed number of samples. The censoring policy in this case works based on a lower threshold, $\lambda_{1}$ and an upper threshold, $\lambda_{2}$. The decision is only being made if the accumulated energy is not in $(\lambda_{1},\lambda_{2})$. For this approach, it is shown that a single-threshold censoring policy is optimal in terms of energy consumption {for both the OR and AND rule}. Moreover, a solution of the underlying problem is given for the OR and AND rule.
\end{itemize}

\subsection{Related work to censoring}

Censoring has been thoroughly investigated in wireless sensor networks and cognitive radios \cite{AVJ,AVJ2,AVJ3,RWB,SZL,MPL3}. It has been shown that censoring is very effective in terms of energy efficiency. In the early works, \cite{AVJ,AVJ2,AVJ3,RWB}, the design of censoring parameters including lower and upper thresholds has been considered and mainly two problem formulations have been studied. In the Neyman-Pearson (NP) case, the miss-detection probability is minimized subject to a constraint on the probability of false alarm and average network energy consumption \cite{AVJ2,AVJ3,RWB}. In the Bayesian case, on the other hand, the detection error probability is minimized subject to a constraint on the average network energy consumption. 
Censoring for cognitive radios is considered in \cite{SZL}, \cite{MPL3}. In \cite{SZL}, a censoring rule similar to the one in this paper is considered in order to limit the bandwidth occupancy of the cognitive radio network. Our fixed sample size censoring scheme is different in two ways. First, in \cite{SZL}, { only the OR rule is considered and} the FC makes no decision in case it does not receive any decision from the cognitive radios which is ambiguous, since the FC has to make a final decision, while in our paper, {the FC reports the absence (for the OR rule) or the presence (for the AND rule) of the primary user, if no local decision is received at the FC}. Second, we give a clear optimization problem and expression for the solution while this is not presented in \cite{SZL}. 
{A combined sleeping and censoring scheme is considered in \cite{MPL3}. The censoring scheme in this paper is different in some ways. The optimization problem in the current paper is defined as the minimization of the maximum {average energy consumption per sensor} while in \cite{MPL3}, the total network energy consumption is minimized. For low-power radios, the problem in this paper makes more sense since the energy of individual radios is generally limited. In this paper, the received SNRs by the cognitive radios are assumed to be different while in \cite{MPL3}, the SNRs are the same. Finally note that the sleeping policy of \cite{MPL3} can be easily incorporated in our proposed censored truncated sequential sensing leading to even higher energy savings.} 

\subsection{Related work to sequential sensing}
Sequential detection as an approach to reduce the average number of sensors required to reach a decision is also studied comprehensively during the past decades \cite{VV,VBP,H, BS,AMM,XZ}.
In \cite{VV}, \cite{VBP}, each sensor collects a sequence of observations, constructs a summary message and passes it on to the FC and all other sensors. A Bayesian problem formulation comprising the minimization of the average error detection probability and sampling time cost over all admissible decision policies at the FC and all possible local decision functions at each sensor is then considered to determine the optimal stopping and decision rule. Further, algorithms to solve the optimization problem for both infinite and finite horizon are given. 
In \cite{H}, an infinite horizon sequential detection scheme based on the sequential probability ratio test (SPRT) at both the sensors and the FC is considered. Wald's analysis of error probability, \cite{W}, is employed to determine the thresholds at the sensors and the FC. 
A combination of sequential detection and censoring is considered in \cite{BS}. Each sensor computes the LLR of the received sample and sends it to the FC, if it is deemed to be in a certain region. The FC then collects the received LLRs and as soon as their sum is larger than an upper threshold or smaller than a lower threshold, the decision is made and the sensors can stop sensing. The LLRs are transmitted in such a way that the larger LLRs are sent sooner. It is shown that the number of transmissions considerably  reduces and particularly when the {transmission} energy is high, this approach performs very well. However, our paper employs a hard fusion scheme at the FC, our sequential scheme is finite horizon, and further a clear optimization problem is given to optimize the energy consumption. { Since we employ the OR (or the AND) rule in our paper, the FC can decide for the presence (or absence) of the primary user by only receiving a single one (or zero). Hence, ordered transmission can be easily incorporated in our paper by stopping the sensing and transmission procedure as soon as one cognitive radio sends a one (or zero) to the FC.} \cite{AMM} proposes a sequential censoring scheme where an SPRT is employed by the FC and soft or hard local decisions are sent to the FC according to a censoring policy. It is depicted that the number of transmissions decreases but on the other hand the average sample number (ASN) increases. Therefore, \cite{AMM} ignores {the effect of sensing on the energy consumption} and focuses only on the transmission energy which for current low-power radios is comparable to the sensing energy. 
A truncated sequential sensing technique is employed in \cite{XZ} to reduce the sensing time of a cognitive radio system. The thresholds are determined such that a certain probability of false alarm and detection are obtained. In this paper, we are employing a similar technique, except that in \cite{XZ}, after the truncation point, a single threshold scheme is used to make a final decision, while in our paper, the sensor decision is censored if no decision is made before the truncation point. Further, \cite{XZ} considers a single sensor detection scheme while we employ a distributed cooperative sensing system and finally, in our paper an explicit optimization problem is given to find the sensing parameters. 

The remainder of the paper is organized as follows. In Section~\ref{censprob}, the fixed size censoring scheme {for the OR rule} is described, including the optimization problem and the algorithm to solve it. The sequential censoring scheme {for the OR rule} is presented in Section~\ref{seqprob}. Analytical expressions for the underlying system parameters are derived and the optimization problem is analyzed. {In Section~\ref{and}, the censoring and sequential censoring schemes are presented and analyzed for the AND rule}. We discuss some numerical results in Section~\ref{numres}. Conclusions and ideas for further work are finally posed in Section~\ref{conc}.
  
\section{Fixed Size Censoring Problem Formulation}\label{censprob}

\begin{figure}[tp]
\centering {\includegraphics[width=3.4in, height=1.6in]{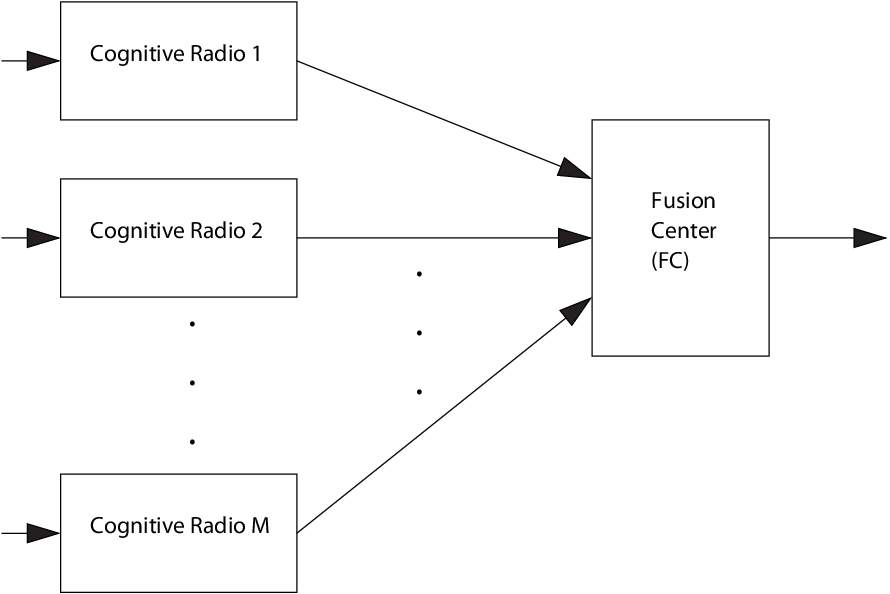}}
\caption{Distributed spectrum sensing configuration} \label{scheme}
\end{figure}

A fixed size censoring scheme is discussed in this section as a benchmark for the main contribution of the paper in Section \ref{seqprob}, which studies a combination of sequential sensing and censoring. A network of $M$ cognitive radios is considered under a cooperative spectrum sensing scheme. A parallel detection configuration is employed as shown in Fig. \ref{scheme}. Each cognitive radio senses the spectrum and makes a local decision about the presence or absence of the primary user and informs the FC by employing a censoring policy. The final decision is then made at the FC by employing the OR rule. {The AND rule will be discussed in Section~\ref{and}}. Denoting $r_{ij}$ to be the $i$-th sample received at the $j$-th cognitive radio, each radio solves a binary hypothesis testing problem as follows
\begin{eqnarray}\label{hyp}
{\cal H}_{0}&:&~r_{ij}=w_{ij},~i=1,...,N,~j=1,...,M\nonumber\\
{\cal H}_{1}&:&~r_{ij}=h_{ij}s_{i}+w_{ij},~i=1,...,N,~j=1,...,M
\end{eqnarray}
where $w_{ij}$ is additive white Gaussian noise with zero mean and variance $\sigma^{2}_{w}$. $h_{ij}$ and $s_{i}$ are the channel gain between the primary user and the $j$-th cognitive radio and the transmitted primary user signal, respectively. {We assume two models for $h_{ij}$ and $s_{i}$. In the first model, $s_{i}$ is assumed to be white Gaussian with zero mean and variance $\sigma^{2}_{s}$, and $h_{ij}$ is assumed constant during each sensing period and thus $h_{ij}=h_{j},~i=1,\dots,N$. In the second model, $s_{i}$ is assumed to be deterministic and constant modulus $|s_{i}|=s,~i=1,\dots,N,~j=1,\dots,M$ and $h_{ij}$ is an i.i.d. Gaussian random process with zero mean and variance $\sigma^{2}_{hj}$. { Note that the second model actually represents a fast fading scenario.} Although each model requires a different type of channel estimation, since the received signal is still a zero mean Gaussian random process with some variance, namely $\sigma_{j}^{2}=h_{j}\sigma^{2}_{s}+\sigma^{2}_{w}$ for the former model and $\sigma_{j}^{2}=s\sigma^{2}_{hj}+\sigma^{2}_{w}$ for the latter model, the analyses which are given in the following sections are valid for both models. The SNR of the received primary user signal at the $j$-th cognitive radio is $\gamma_{j}=|h_{j}|^{2}\sigma^{2}_{s}/\sigma^{2}_w$ under the first model and $\gamma_{j}=s^{2}\sigma^{2}_{hj}/\sigma^{2}_w$ under the second model.} Furthermore, $h_{ij}{s_{i}}$ and ${w_{ij}}$ are assumed statistically independent.

An energy detector is employed by each cognitive sensor which calculates the accumulated energy over $N$ observation samples. {Note that under our system model parameters, the energy detector is equivalent to the optimal LLR detector \cite{Kay}}. The received energy collected over the $N$ observation samples at the $j$-th radio is given by
\begin{equation} \label{edm}
{\cal E}_{j}=\sum^{N}_{i=1}\frac{|r_{ij}|^{2}}{{\sigma^{2}_{w}}}.
\end{equation}

When the accumulated energy of the observation samples is calculated, a censoring policy is employed at each radio where the local decisions are sent to the FC only if they are deemed to be informative \cite{MPL3}. Censoring
thresholds $\lambda_1$ and $\lambda_2$ are applied at each of the radios, where the range
$\lambda_1<{\cal E}_{j}<\lambda_2$ is called the censoring region. At the $j$-th radio, the local censoring
decision rule is given by
\begin{equation} \label{censrule}
\left\{
\begin{array}{lr}
\mbox{send 1, declaring ${\cal H}_1$} & \text{if } {\cal E}_{j}\ge{\lambda_{2}},\\
\mbox{no decision} & \text{if}~
\lambda_{1}<{\cal E}_{j}<\lambda_{2},\\
\mbox{send 0, declaring ${\cal H}_0$} & \text{if }
{\cal E}_{j}\le{\lambda_{1}}.
\end{array} \right.
\end{equation}

It is well known \cite{Kay} that under such a model, ${\cal E}_{j}$ follows a central chi-square distribution with $2N$ degrees of
freedom under ${\cal H}_0$ and ${\cal H}_1$. 
Therefore, the local probabilities of false alarm
and detection can be respectively written as
\begin{align} 
P_{fj}&=Pr({\cal E}_{j}\ge{\lambda_{2}}|{\cal
H}_0)=\frac{\Gamma(N,\frac{\lambda_{2}}{2})}{\Gamma(N)},\label{pfcensexp}\\
P_{dj}&=Pr({\cal E}_{j}\ge{\lambda_{2}}|{\cal
H}_1)=\frac{\Gamma(N,\frac{\lambda_{2}}{2(1+\gamma_{j})})}{\Gamma(N)},\label{pdcensexp}
\end{align}
where $\Gamma(a,x)$ is the incomplete gamma function given by
$\Gamma(a,x)=\int^{\infty}_{x}t^{a-1}e^{-t}dt$, with
$\Gamma(a,0)=\Gamma(a)$.

Denoting $C_{sj}$ and $C_{ti}$ to be the energy consumed by the $j$-th radio in sensing per sample and
transmission per bit, respectively, the average energy consumed for
distributed sensing per user is given by,
\begin{equation} \label{avecost}
C_{j}= NC_{sj}+(1-\rho_{j})C_{tj},
\end{equation}
where $\rho_{j}=Pr(\lambda_1<{\cal E}_{j}<\lambda_2)$ is denoted to be the average censoring rate. Note that $C_{sj}$ is fixed and only depends on the sampling rate and power consumption of the sensing module while $C_{tj}$ depends on the distance to the FC at the time of the transmission. Therefore, in this paper, it is assumed that the cognitive radio is aware of its location {and the location of the FC as well as their mutual channel properties} or at least can estimate them. Defining $\pi_{0}=Pr({\cal H}_{0})$, $\pi_{1}=Pr({\cal H}_{1})$, $\delta_{0j}=Pr(\lambda_1<{\cal E}_{j}<\lambda_2|{\cal H}_{0})$ and $\delta_{1j}=Pr(\lambda_1<{\cal E}_{j}<\lambda_2|{\cal H}_{1})$, $\rho_{j}$ is given by
\begin{eqnarray}\label{rhocensexp}
\rho_{j}
=\pi_{0}\delta_{0j}+\pi_{1}\delta_{1j},
\end{eqnarray}
with
\begin{eqnarray}
\delta_{0j}&=&\frac{\Gamma(N,\frac{\lambda_{1}}{2})}{\Gamma(N)}-\frac{\Gamma(N,\frac{\lambda_{2}}{2})}{\Gamma(N)},\label{delta0censexp}\\
\delta_{1j}&=&\frac{\Gamma(N,\frac{\lambda_{1}}{2(1+\gamma_{j})})}{\Gamma(N)}-\frac{\Gamma(N,\frac{\lambda_{2}}{2(1+\gamma_{j})})}{\Gamma(N)}.\label{delta1censexp}
\end{eqnarray}

Denoting $Q^{c}_{\text{F}}$ and $Q^{c}_{\text{D}}$ to be the respective global probability of false alarm and
detection, the target detection performance is then quantified by
$Q^{c}_{\text{F}} \le \alpha$ and $Q^{c}_{\text{D}} \ge \beta$, where $\alpha$ and $\beta$ are
pre-specified detection design parameters. 
Our goal is to determine the optimum censoring
thresholds $\lambda_1$ and $\lambda_2$ such that the maximum {average energy consumption per sensor}, i.e., $\max_{j}{~C_{j}}$, is minimized
subject to the constraints $Q^{c}_{\text{F}} \le \alpha$ and $Q^{c}_{\text{D}} \ge \beta$. Hence, our
optimization problem can be formulated as
\begin{align} \label{opt1}
&\underset{\lambda_1,\lambda_{2}}\min~\max_{j}{~C_{j}}\nonumber\\
 &\text{s.t.}~Q^{c}_{\text{F}}\le{\alpha},~Q^{c}_{\text{D}}\ge{\beta}.
 \end{align}
 
In this section, the FC employs an OR rule to
make the final decision which is denoted by $D_{{FC}}$, i.e., $D_{{FC}}=1$ if the FC receives at
least one local decision declaring 1, else $D_{{FC}}=0$. This way, the global probability of false alarm and detection can be derived as
\begin{align}
&Q^{c}_{\text{F}}=Pr(D_{FC}=1|{\cal H}_{0})=1-\prod^{M}_{j=1}(1-P_{fj}),\label{qfcensexp}\\
&Q^{c}_{\text{D}}=Pr(D_{FC}=1|{\cal H}_{1})=1-\prod^{M}_{j=1}(1-P_{dj}).\label{qdcensexp}
\end{align}Note that since all the cognitive radios employ the same upper threshold $\lambda_{2}$, we can state that $P_{fj}=P_{f}$ defined in (\ref{pfcensexp}). As a result, (\ref{qfcensexp}) becomes
\begin{equation}\label{qfcensexp2}
Q^{c}_{\text{F}}=1-(1-P_{f})^M.
\end{equation}

Since the FC decides about the presence of the primary user only by receiving 1s (receiving no decision from all the sensors is considered as absence of the primary user) and the sensing time does not depend on $\lambda_{1}$, it is a waste of energy to send zeros to the FC and thus, the optimal solution of (\ref{opt1}) is obtained by $\lambda_{1}=0$. Note that this is only the case for fixed-size censoring, because the energy consumption of each sensor only varies by the transmission energy while the sensing energy is constant. 
{This way (\ref{delta0censexp}) and (\ref{delta1censexp}) can be simplified to $\delta_{0j}=1-P_{f}$ and $\delta_{1j}=1-P_{dj}$, and we only need to derive the optimal $\lambda_{2}$. Since there is a one-to-one relationship between $P_{f}$ and $\lambda_{2}$, by finding the optimal $P_{f}$, $\lambda_{2}$ can also be easily derived as $\lambda_{2}=2\Gamma^{-1}[N,\Gamma(N)P_f]$ (where $\Gamma^{-1}$ is defined over the second argument). Considering this result and defining $Q^{c}_{\text{D}}=H(P_{f})$, the optimal solution of (\ref{opt1}) is given by $P_{f}=H^{-1}(\beta)$ as is shown in Appendix~\ref{optpf}.} 

In the following section, a combination of censoring and sequential sensing approaches is presented which optimizes both the sensing and the transmission energy.

\section{Sequential Censoring Problem Formulation}\label{seqprob}
\subsection{System Model}\label{sysmod}

Unlike Section \ref{censprob}, where each user collects a specific number of samples, in this section, each cognitive radio sequentially senses the spectrum and upon reaching a decision about the presence or absence of the primary user, it sends the result to the FC by employing a censoring policy as introduced in Section \ref{censprob}. The final decision is then made at the FC by employing the OR rule. Here, a censored truncated sequential sensing scheme is employed where each cognitive radio carries on sensing until it reaches a decision while not passing a limit of $N$ samples. We define $\zeta_{nj}=\sum^{n}_{i=1}|r_{ij}|^{2}/\sigma^{2}_{w}=\sum^{n}_{i=1}x_{ij}$ and $a_{i}=0,~i=1,\dots,p$, $a_{i}=\bar{a}+i\bar{\Lambda},~i=p+1,...,N$ and $b_{i}=\bar{b}+i\bar{\Lambda},~i=1,...,N$, where $\bar{a}=a/\sigma^{2}_{w}$, $\bar{b}=b/\sigma^{2}_{w}$, $1<\bar{\Lambda}<1+\gamma_{j}$ is a predetermined constant, $a<0$, $b>0$ and $p=\lfloor{-a/\sigma^{2}_{w}\bar{\Lambda}}\rfloor$ \cite{XZ}. We assume that the SNR $\gamma_{j}$ is known or can be estimated. This way, the local decision rule in order to make a final decision is as follows 
\begin{equation} \label{censrule2}
\left\{
\begin{array}{lr}
\mbox{send 1, declaring ${\cal H}_1$} & \text{if } \zeta_{nj}\ge{b_{n}}~ \text{and}~ n\in[1,N],\\
\mbox{continue sensing} &\text{if}~\zeta_{nj}\in(a_{n},b_{n})~\text{and}~n\in[1,N),\\
\mbox{no decision} & \text{if}~
\zeta_{nj}\in(a_{n},b_{n})~\text{and}~n=N,\\
\mbox{send 0, declaring ${\cal H}_0$} & \text{if }
\zeta_{nj}\le{a_{n}}~\text{and}~ n\in[1,N].
\end{array} \right.
\end{equation}

The probability density function of $x_{ij}=|r_{ij}|^{2}/\sigma^{2}_{w}$ under ${\cal H}_{0}$ and ${\cal H}_{1}$ is a chi-square distribution with $2n$ degrees of freedom. Thus, $x_{ij}$ becomes exponentially distributed under both ${\cal H}_{0}$ and ${\cal H}_{1}$. Henceforth, we obtain
\begin{eqnarray}
Pr(x_{ij}|{{\cal H}_{0}})&=&\frac{1}{2}e^{-x_{ij}/2}I_{\{x_{ij}\ge{0}\}},\\
Pr(x_{ij}|{{\cal H}_{1}})&=&\frac{1}{2(1+\gamma_{j})}e^{-x_{ij}/2(1+\gamma_{j})}I_{\{x_{ij}\ge{0}\}},
\end{eqnarray}
where $I_{\{x_{ij}\ge{0}\}}$ is the indicator function.

Defining $\zeta_{0j}=0$, the local probability of false alarm at the $j$-th cognitive radio, $P_{fj}$, can be written as
\begin{eqnarray}\label{pf}
P_{fj}&=&\sum^{N}_{n=1}Pr(\zeta_{0j}\in(a_{0},b_{0}),...,\zeta_{n-1j}\in(a_{n-1},b_{n-1}),\zeta_{nj}\ge{b_{n}}|{\cal H}_{0}),
\end{eqnarray}
whereas the local probability of detection, $P_{dj}$, is obtained as follows
\begin{eqnarray}\label{pd}
P_{dj}&=&\sum^{N}_{n=1}Pr(\zeta_{0j}\in(a_{0},b_{0}),...,\zeta_{n-1j}\in(a_{n-1},b_{n-1}),\zeta_{nj}\ge{b_{n}}|{\cal H}_{1}).
\end{eqnarray}

Denoting $\rho_{j}$ to be the average censoring rate at the $j$-th cognitive radio, and $\delta_{0j}$ and $\delta_{1j}$ to be the respective average censoring rate under ${\cal H}_{0}$ and ${\cal H}_{1}$, we have
\begin{eqnarray}\label{censrate}
\rho_{j}
=\pi_{0}\delta_{0j}+\pi_{1}\delta_{1j},
\end{eqnarray}
where
\begin{align}
&\delta_{0j}=Pr(\zeta_{1j}\in(a_{1},b_{1}),...,\zeta_{Nj}\in(a_{N},b_{N})|{\cal H}_{0}),\label{delta0}\\
&\delta_{1j}=Pr(\zeta_{1j}\in(a_{1},b_{1}),...,\zeta_{Nj}\in(a_{N},b_{N})|{\cal H}_{1}).\label{delta1}
\end{align}

The other parameter that is important in any sequential detection scheme is the average sample number (ASN) required to reach a decision. Denoting $N_{j}$ to be a random variable representing the number of samples required to announce the presence or absence of the primary user, the ASN for the $j$-th cognitive radio, denoted as $\bar{N}_{j}$=$E(N_{j})$, can be defined as 
\begin{eqnarray}\label{ASN}
\bar{N}_{j}=\pi_{0}E(N_{j}|{\cal H}_{0})+\pi_{1}E(N_{j}|{\cal H}_{1}),
\end{eqnarray}
where
\begin{eqnarray}\label{ensh0}
E(N_{j}|{\cal H}_{0})&=&\sum^{N}_{n=1}nPr(N_{j}=n|{\cal H}_{0})\nonumber\\
&=&\sum^{N-1}_{n=1}n[Pr(\zeta_{0j}\in{(a_{0},b_{0})},...,\zeta_{n-1j}\in{(a_{n-1},b_{n-1})}|{\cal H}_{0})\nonumber\\
&-&Pr(\zeta_{0j}\in{(a_{0},b_{0})},...,\zeta_{nj}\in{(a_{n},b_{n})}|{\cal H}_{0})]\nonumber\\
&+&NPr(\zeta_{0j}\in{(a_{0},b_{0})},...,\zeta_{N-1j}\in{(a_{N-1},b_{N-1})}|{\cal H}_{0}),
\end{eqnarray}
and
\begin{eqnarray}\label{ensh1}
E(N_{j}|{\cal H}_{1})&=&\sum^{N}_{n=1}nPr(N_{j}=n|{\cal H}_{1})\nonumber\\
&=&\sum^{N-1}_{n=1}n[Pr(\zeta_{0j}\in{(a_{0},b_{0})},...,\zeta_{nj}\in{(a_{n-1},b_{n-1})}|{\cal H}_{1})\nonumber\\
&-&Pr(\zeta_{0j}\in{(a_{0},b_{0})},...,\zeta_{nj}\in{(a_{n},b_{n})}|{\cal H}_{1})]\nonumber\\
&+&NPr(\zeta_{0j}\in{(a_{0},b_{0})},...,\zeta_{N-1j}\in{(a_{N-1},b_{N-1})}|{\cal H}_{1}).
\end{eqnarray}

Denoting again $C_{sj}$ to be the sensing energy of one sample and $C_{tj}$ to be the transmission energy of a decision bit at the $j$-th cognitive radio, the total average energy consumption at the $j$-th cognitive radio now becomes
\begin{equation}\label{cost}
C_{j}=\bar{N}_{j}C_{sj}+(1-\rho_{j})C_{tj}.
\end{equation}

Denoting $Q^{cs}_{\text{F}}$ and $Q^{cs}_{\text{D}}$ to be the respective global probabilities of false alarm and detection for the censored truncated sequential approach, we define our problem as the minimization of the maximum average energy consumption per sensor subject to a constraint on the global probabilities of false alarm and detection as follows
\begin{align}\label{probform}
&\min_{\bar{a},\bar{b}}{~\max_{j}{~C_{j}}}\nonumber\\
&\text{s.t.}~Q^{cs}_{\text{F}}\le{\alpha},~Q^{cs}_{\text{D}}\ge{\beta}.
\end{align}

As in (\ref{qfcensexp}) and (\ref{qdcensexp}), under the OR rule that is assumed in this section, the global probability of false alarm is
\begin{equation}\label{qf}
Q^{cs}_{\text{F}}=Pr(D_{\text{FC}}=1|{\cal H}_{0})=1-\prod^{M}_{j=1}(1-P_{fj}),
\end{equation}
and the global probability of detection is
\begin{equation}\label{qd}
Q^{cs}_{\text{D}}=Pr(D_{\text{FC}}=1|{\cal H}_{1})=1-\prod^{M}_{j=1}(1-P_{dj}).
\end{equation}
Note that since $P_{f1}=\dots=P_{fM}$, it is again assumed that $P_{fj}=P_f$ in this section.

In the following subsection, analytical expressions for the probability of false alarm and detection as well as the censoring rate and ASN are extracted.

\subsection{Parameter and Problem Analysis}\label{prob}

Looking at (\ref{pf}), (\ref{pd}), (\ref{censrate}) and (\ref{ASN}), we can see that the joint probability distribution function of $p(\zeta_{1j},...,\zeta_{nj})$ is the foundation of all the equations. Since $x_{ij}=\zeta_{ij}-\zeta_{i-1j}$ for $i=1,...,N$, we have,
\begin{eqnarray}\label{jointprob}
p(\zeta_{1j},...,\zeta_{nj})
=p(x_{nj})p(x_{n-1j})...p(x_{1j}).
\end{eqnarray}

Therefore, the joint probability distribution function under ${\cal H}_{0}$ and ${\cal H}_{1}$ becomes
\begin{eqnarray}
p(\zeta_{1j},...,\zeta_{nj}|{\cal H}_{0})&=&\frac{1}{2^{n}}e^{-\zeta_{nj}/2}I_{\{0\le{{\zeta_{1j}}\le{\zeta_{2j}}...\le{\zeta_{nj}}\}}},\\
p(\zeta_{1j},...,\zeta_{nj}|{\cal H}_{1})&=&\frac{1}{[2(1+\gamma_{j})]^{n}}e^{-\zeta_{nj}/2(1+\gamma_{j})}I_{\{0\le{\zeta_{1j}}\le{\zeta_{2j}}...\le{\zeta_{nj}}\}},
\end{eqnarray}
where $I_{\{0\le{\zeta_{1j}}\le{\zeta_{2j}}...\le{\zeta_{nj}}\}}$ is again the indicator function.

The derivation of the local probability of false alarm and the ASN under ${\cal H}_{0}$ in this work are similar to the ones considered in \cite{XZ} and \cite{WK}. The difference is that in \cite{XZ}, if the cognitive radio does not reach a decision after $N$ samples, it employs a single threshold decision policy to give a final decision about the presence or absence of the cognitive radio, while in our work, no decision is sent in case none of the upper and lower thresholds are crossed. Hence, to avoid introducing a cumbersome detailed derivation of each parameter, we can use the results in \cite{XZ} for our analysis with a small modification. However, note that the problem formulation in this work is essentially different from the one in \cite{XZ}. Further, since in our work the distribution of $x_{ij}$ under ${\cal H}_{1}$ is exponential like the one under ${\cal H}_{0}$, unlike \cite{XZ}, we can also use the same approach to derive analytical expressions for the local probability of detection, the ASN under ${\cal H}_{1}$, and the censoring rate.

Denoting $E_{n}$ to be the event where $a_{i}<\zeta_{ij}<b_{i},~i=1,...,n-1$ and $\zeta_{nj}\ge{b_{n}}$, (\ref{pf}) becomes
\begin{equation}\label{pfseqexp}
P_{fj}=\sum^{N}_{n=1}Pr(E_{n}|{\cal H}_{0}).
\end{equation}
where the analytical expression for $Pr(E_{n}|{\cal H}_{0})$ is derived in Appendix~\ref{prenh0}.

Similarly for the local probability of detection, we have
\begin{equation}\label{pdseqexp}
P_{dj}=\sum^{N}_{n=1}Pr(E_{n}|{\cal H}_{1}),
\end{equation}
where the analytical expression for $Pr(E_{n}|{\cal H}_{1})$ is derived in Appendix~\ref{prenh1}. 

Defining $R_{nj}=\{\zeta_{ij}|\zeta_{ij}\in(a_{i},b_{i}),~i=1,...,n\}$, $Pr(R_{nj}|{\cal H}_{0})$ and $Pr(R_{nj}|{\cal H}_{1})$ are obtained as follows
\begin{equation}
Pr(R_{nj}|{\cal H}_{0})=\frac{1}{2^{n}}J^{(n)}_{a_{n},b_{n}}(1/2),~n=1,...,N,
\end{equation}
\begin{equation}
Pr(R_{nj}|{\cal H}_{1})=\frac{1}{[2(1+\gamma_{j})]^{n}}J^{(n)}_{a_{n},b_{n}}(1/2(1+\gamma_{j})),~n=1,...,N,
\end{equation}
where $J^{(n)}_{a_{n},b_{n}}(\theta)$ is presented in Appendix~\ref{j} and (\ref{ensh0}) and (\ref{ensh1}) become
\begin{equation}\label{ensh0exp}
E(N_{j}|{\cal H}_{0})=\sum^{N-1}_{n=1}n(Pr(R_{n-1j}|{\cal H}_{0})-Pr(R_{nj}|{\cal H}_{0}))+NPr(R_{N-1j}|{\cal H}_{0})=1+\sum^{N-1}_{n=1}Pr(R_{nj}|{\cal H}_{0}),
\end{equation}
\begin{equation}\label{ensh1exp}
E(N_{j}|{\cal H}_{1})=\sum^{N}_{n=1}n(Pr(R_{n-1j}|{\cal H}_{1})-Pr(R_{nj}|{\cal H}_{1}))+NPr(R_{N-1j}|{\cal H}_{1})=1+\sum^{N-1}_{n=1}Pr(R_{nj}|{\cal H}_{1}).
\end{equation}

With (\ref{ensh0exp}) and (\ref{ensh1exp}), we can calculate (\ref{ASN}). This way, (\ref{delta0}) and (\ref{delta1}) can be derived as follows
\begin{equation}\label{delta0exp}
\delta_{0j}=Pr(R_{Nj}|{\cal H}_{0})=\frac{1}{2^{N}}J^{(N)}_{a_{N},b_{N}}(1/2),
\end{equation}
\begin{equation}\label{delta1exp}
\delta_{1j}=Pr(R_{Nj}|{\cal H}_{1})=\frac{1}{[2(1+\gamma_{j})]^{N}}J^{(N)}_{a_{N},b_{N}}(1/2(1+\gamma_{j})).
\end{equation}

We can show that the problem (\ref{probform}) is not convex. Therefore, the standard systematic optimization algorithms do not give the global optimum for $\bar{a}$ and $\bar{b}$. However, as is shown in the following lines, $\bar{a}$ and $\bar{b}$ are bounded and therefore, a two-dimensional exhaustive search is possible to find the global optimum. First of all, we have $a<0$ and $\bar{a}<0$. On the other hand, if $\bar{a}$ has to play a role in the sensing system, at least one $a_{N}$ should be positive, i.e., $a_{N}=\bar{a}+N\Delta\ge 0$ which gives $\bar{a}\ge -N\Delta$. Hence, we obtain $-N\Delta\le{\bar{a}}<0$. Furthermore, defining $Q^{cs}_{\text{F}}=\mathcal{F}(\bar{a},\bar{b})$ and $Q^{cs}_{\text{D}}=\mathcal{G}(\bar{a},\bar{b})$, for a given $\bar{a}$, it is easy to show that $\mathcal{G}^{-1}(\bar{a},\beta)\le{\bar{b}}\le{\mathcal{F}^{-1}(\bar{a},\alpha)}$ (where $\mathcal{F}^{-1}$ and $\mathcal{G}^{-1}$ are defined over the second argument). 

Before introducing a suboptimal problem, the following theorem is presented.

\emph{Theorem~1}. For a given local probability of detection and false alarm ($P_{d}$ and $P_{f}$) and $N$, the censoring rate of the optimal censored truncated sequential sensing ($\rho^{cs}$) is less than the one of the censoring scheme ($\rho^{c}$).

\textbf{Proof}. The proof is provided in Appendix~\ref{thr1}.

We should note that, in censored truncated sequential sensing, a large amount of energy is to be saved on sensing. Therefore, as is shown in Section~\ref{numres}, as the sensing energy of each sensor increases, censored truncated sequential sensing outperforms censoring in terms of energy efficiency. However, in case that the transmission energy is much higher than the sensing energy, it may happen that censoring outperforms censored truncated sequential sensing, because of a higher censoring rate ($\rho^{cs}>\rho^{c}$). Hence, one corollary of Theorem~1 is that although the optimal solution of (\ref{opt1}) for a specific $N$, i.e., $P_{d}=1-(1-\beta)^{1/M}$ and $P_{f}=H^{-1}(\beta)$, is in the feasible set of (\ref{probform}) for a resulting ASN less than $N$, it does not necessarily guarantee that the resulting {average energy consumption per sensor} of the censored truncated sequential sensing approach is less than the one of the censoring scheme, particularly when the transmission energy is much higher than the sensing energy per sample.

Solving (\ref{probform}) is complex in terms of the number of computations, and thus a two-dimensional exhaustive search is not always a good solution. Therefore, in order to reach a good solution in a reasonable time, we set $a<{-N\Delta}$ in order to obtain $a_{1}=\dots=a_{N}={0}$. This way, we can relax one of the arguments of (\ref{probform}) and only solve the following suboptimal problem
\begin{align}\label{suboptprob}
&\min_{\bar{b}}~\max_{j}{~C_{j}}\nonumber\\
&\text{s.t.}~Q^{cs}_{\text{F}}\le{\alpha},~Q^{cs}_{\text{D}}\ge{\beta}.
\end{align}
Note that unlike Section \ref{censprob}, here the zero lower threshold is not necessarily optimal. The reason is that although the maximum censoring rate is achieved with the lowest $\bar{a}$, the minimum ASN is achieved with the highest $\bar{a}$, and thus there is an inherent trade-off between a high censoring rate and a low ASN and a zero $a_{i}$ is not necessarily the optimal solution. Since the analytical expressions provided earlier are very complex, we now try to provide a new set of analytical expressions for different parameters based on the fact that $a_{1}=\dots={a_{N}}={0}$.

To find an analytical expression for $P_{fj}$, we can derive $A(n)$ for the new paradigm as follows
\begin{equation}
A(n)=\underset{\Gamma_{n}}{\int...\int}I_{\{0\le{{\zeta_{1j}}\le{\zeta_{2j}}...\le{\zeta_{n-1j}}\}}d\zeta_{1j}...d\zeta_{n-1j}}.
\end{equation}

Since $0\le{{\zeta_{1j}}\le{\zeta_{2j}}...\le{\zeta_{n-1j}}}$ and $a_{1}=\dots=a_{N}={0}$, the lower bound for each integral is $\zeta_{i-1}$ and the upper bound is $b_{i}$, where $i=1,...,n-1$. Thus we obtain
\begin{equation}
A(n)=\int^{b_{1}}_{\zeta_{0j}}\int^{b_{2}}_{\zeta_{1j}}...\int^{b_{n-1}}_{\zeta_{n-2j}}d\zeta_{1j}d\zeta_{2j}...d\zeta_{n-1j},
\end{equation}
which according to \cite{WK} is
\begin{equation}\label{anexp}
A(n)=\frac{b_{1}b^{n-2}_{n}}{(n-1)!},~n=1,...,N.
\end{equation}
Hence, we have
\begin{equation}\label{pfexp2}
P_{fj}=\sum^{N}_{n=1}p_{n}A(n),
\end{equation}
and $p_{n}=\frac{e^{-b_{n}/2}}{2^{n-1}}$. Similarly, for $P_{dj}$, we obtain
\begin{eqnarray}
B(n)&=&\int^{b_{1}}_{\zeta_{0j}}\int^{b_{2}}_{\zeta_{1j}}...\int^{b_{n-1}}_{\zeta_{n-2j}}d\zeta_{1j}d\zeta_{2j}...d\zeta_{n-1j}\nonumber\\
&=&\frac{b_{1}b^{n-2}_{n}}{(n-1)!},~n=1,...,N,
\end{eqnarray}
and thus
\begin{equation}\label{pdexp2}
P_{dj}=\sum^{N}_{n=1}q_{n}B(n),
\end{equation}
where $q_{n}=\frac{e^{-b_{n}/2(1+\gamma_{j})}}{[2(1+\gamma_{j})]^{n-1}}$. Furthermore, we note that for $a_{1}=\dots={a_{N}}={0}$, $A(n)=B(n)=\frac{b_{1}b^{n-2}_{n}}{(n-1)!},~n=1,...,N$.

It is easy to see that $R_{nj}$ occurs under ${\cal H}_{0}$, if no false alarm happens until the $n$-th sample. Therefore, the analytical expression for $Pr(R_{nj}|{\cal H}_{0})$ is given by
\begin{eqnarray}\label{rnh0}
Pr(R_{nj}|{\cal H}_{0})=1-\sum^{n}_{i=1}p_{i}A(i),
\end{eqnarray}
and in the same way, for $Pr(R_{nj}|{\cal H}_{1})$, we obtain
\begin{eqnarray}\label{rnh1}
Pr(R_{nj}|{\cal H}_{1})=1-\sum^{n}_{i=1}q_{i}A(i).
\end{eqnarray}

Putting (\ref{rnh0}) and (\ref{rnh1}) in (\ref{ensh0exp}) and (\ref{ensh1exp}), we obtain
\begin{equation}\label{ensh0exp2}
E(N_{j}|{\cal H}_{0})=1+\sum^{N-1}_{n=1}\bigg\{1-\sum^{n}_{i=1}p_{i}A(i)\bigg\},
\end{equation}
\begin{equation}\label{ensh1exp2}
E(N_{j}|{\cal H}_{1})=1+\sum^{N-1}_{n=1}\bigg\{1-\sum^{n}_{i=1}q_{i}A(i)\bigg\},
\end{equation}
and inserting (\ref{ensh0exp2}) and (\ref{ensh1exp2}) in (\ref{ASN}), we obtain
\begin{equation}\label{asnexp}
\bar{N}_{j}=\pi_{0}\Bigg(1+\sum^{N-1}_{n=1}\bigg\{1-\sum^{n}_{i=1}p_{i}A(i)\bigg\}\Bigg)+\pi_{1}\Bigg(1+\sum^{N-1}_{n=1}\bigg\{1-\sum^{n}_{i=1}q_{i}A(i)\bigg\}\Bigg).
\end{equation}

Finally, from (\ref{rnh0}) and (\ref{rnh1}), the censoring rate can be easily obtained as
\begin{equation}\label{censexp}
\rho_{j}=\pi_{0}\bigg(1-\sum^{N}_{i=1}p_{i}A(i)\bigg)+\pi_{1}\bigg(1-\sum^{N}_{i=1}q_{i}A(i)\bigg).
\end{equation}

Having the analytical expressions for (\ref{suboptprob}), we can easily find the optimal maximum {average energy consumption per sensor} by a line search over $\bar{b}$. Similar to the censoring problem formulation, here the sensing threshold is also bounded by ${Q^{cs}_{\text{F}}}^{-1}(\alpha)\le{\bar{b}}\le{{Q^{cs}_{\text{D}}}^{-1}(\beta)}$. As we will see in Section \ref{numres}, censored truncated sequential sensing performs better than censored spectrum sensing in terms of energy efficiency for low-power radios.

{
\section{Extension to the AND rule}\label{and}

So far, we have mainly focused on the OR rule. However, another rule which is also simple in terms of implementation is the AND rule. According to the AND rule, $D_{FC}=0$, if at least one cognitive radio reports a zero, else $D_{FC}=1$. This way the global probabilities of false alarm and detection, can be written respectively as
\begin{equation}\label{qfand}
Q^{c}_{\text{F,AND}}=Q^{cs}_{\text{F,AND}}=Pr(D_{FC}=1|{\cal H}_{0})=\prod^{M}_{j=1}(\delta_{0j}+P_{fj}),
\end{equation}
\begin{equation}\label{qdand}
Q^{c}_{\text{D,AND}}=Q^{cs}_{\text{D,AND}}=Pr(D_{FC}=1|{\cal H}_{1})=\prod^{M}_{j=1}(\delta_{1j}+P_{dj}).
\end{equation}
Note that (\ref{qfand}) and (\ref{qdand}) hold for both the sequential censoring and censoring schemes. Similar to the case for the OR rule, the problem is defined so as to minimize the maximum {average energy consumption per sensor} subject to a lower bound on the global probability of detection and an upper bound on the global probability of false alarm. In the following two subsections, we are going to analyze the problem for censoring and sequential censoring.

\subsection{AND rule for fixed-sample size censoring}\label{censand}

The optimization problem for the censoring scheme considering the AND rule at the FC, becomes
\begin{align} \label{andopt}
&\underset{\lambda_1,\lambda_{2}}\min~\max_{j}{~C_{j}}\nonumber\\
 &\text{s.t.}~Q^{c}_{\text{F,AND}}\le{\alpha},~Q^{c}_{\text{D,AND}}\ge{\beta}.
 \end{align}
where $C_{j}$ is defined in (\ref{avecost}). Since the FC decides for the absence of the primary user by receiving at least one zero and the fact that the sensing energy per sample is constant, the optimal upper threshold $\lambda_{2}$ is $\lambda_{2}\rightarrow{\infty}$. This way, cognitive radios censor all the results for which ${\cal E}_{j}>\lambda_{1}$, and as a result (\ref{qfand}) and (\ref{qdand}) become
\begin{equation}\label{qfandcens}
Q^{c}_{\text{F,AND}}=Pr(D_{FC}=1|{\cal H}_{0})=\prod^{M}_{j=1}\delta_{0j},
\end{equation}
\begin{equation}\label{qdandcens}
Q^{c}_{\text{D,AND}}=Pr(D_{FC}=1|{\cal H}_{1})=\prod^{M}_{j=1}\delta_{1j}.
\end{equation}
where $\delta_{0j}=Pr({\cal E}_{j}>\lambda_{1}|{\cal H}_{0})$ and $\delta_{1j}=Pr({\cal E}_{j}>\lambda_{1}|{\cal H}_{1})$. Since the thresholds are the same among the cognitive radios, we have $\delta_{01}=\delta_{02}=\dots=\delta_{0M}=\delta_{0}$. Since there is a one-to-one relationship between $\lambda_{1}$ and $\delta_{0}$, by finding the optimal $\delta_{0}$, the optimal $\lambda_{1}$ can be easily derived. As shown in Appendix~\ref{optdelta0}, we can derive the optimal $\delta_{0}$ as $\delta_{0}=\alpha^{1/M}$. This result is very important in the sense that as far as the feasible set of (\ref{andopt}) is not empty, the optimal solution of (\ref{andopt}) is independent from the SNR. Note that the maximum {average energy consumption per sensor} still depends on the SNR via $\delta_{1j}$ and is reducing as the SNR grows.

\subsection{AND rule for censored truncated sequential sensing}

The optimization problem for the censored truncated sequential sensing scheme with the AND rule, becomes
\begin{align}\label{csprobform}
&\min_{\bar{a},\bar{b}}{~\max_{j}{~C_{j}}}\nonumber\\
&\text{s.t.}~Q^{cs}_{\text{F,AND}}\le{\alpha},~Q^{cs}_{\text{D,AND}}\ge{\beta}.
\end{align}
where $C_{j}$ is defined in (\ref{cost}). Similar to the OR rule, we have $-N\Delta\le{\bar{a}}<0$. Defining $Q^{cs}_{\text{F,AND}}={\cal F}_{\text{AND}}(\bar{a},\bar{b})$ and $Q^{cs}_{\text{D,AND}}={\cal G}_{\text{AND}}(\bar{a},\bar{b})$, for a given $\bar{a}$, we can show that $\mathcal{G}_{\text{AND}}^{-1}(\bar{a},\beta)\le{\bar{b}}\le{\mathcal{F}_{\text{AND}}^{-1}(\bar{a},\alpha)}$ (where $\mathcal{F}_{\text{AND}}^{-1}$ and $\mathcal{G}_{\text{AND}}^{-1}$ are defined over the second argument). Therefore, the optimal $\bar{a}$ and $\bar{b}$ can again be derived by a bounded two-dimensional search, in a similar way as for the OR rule.}

\section{Numerical Results}\label{numres}

{A network of cognitive radios is considered for the numerical results. In some of the scenarios, for the sake of simplicity, it is assumed that all the sensors experience the same SNR. This way, it is easier to show how the main performance indicators including the optimal maximum {average energy consumption per sensor}, ASN and censoring rate changes when one of the underlying parameter of the system changes. However, to comply with the general idea of the paper, which is based on different received SNRs by cognitive radios, in other scenarios, the different cognitive radios experience different SNRs. Unless otherwise mentioned, the results are based on the single-threshold strategy for censored truncated sequential sensing in case of the OR rule. }

\begin{figure}[!tp]
\centerline{\subfloat[]{\includegraphics[width=3.2in, height=2.6in]{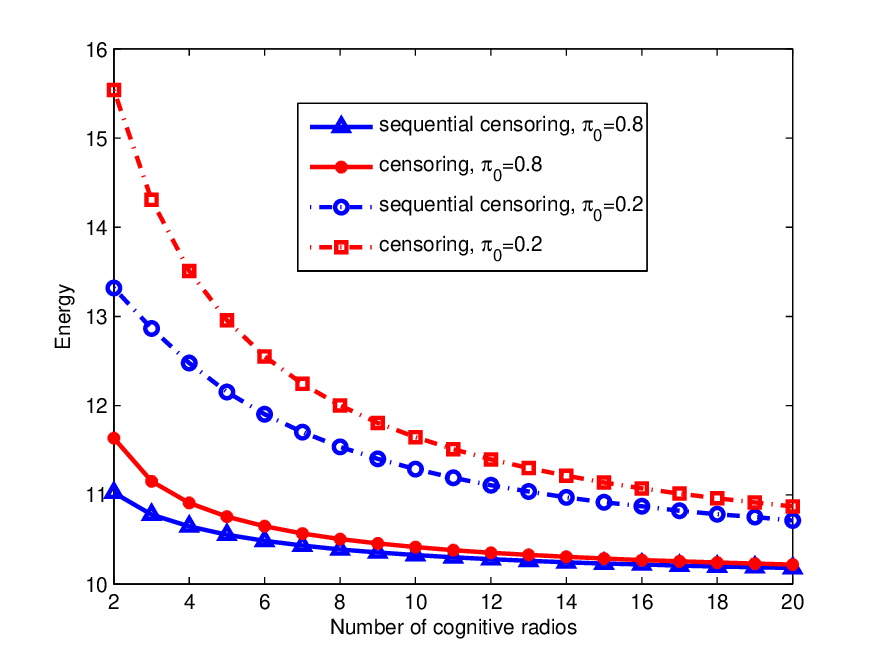}%
\label{em}}}
\hfil
\centerline{\subfloat[]{\includegraphics[width=3.2in, height=2.6in]{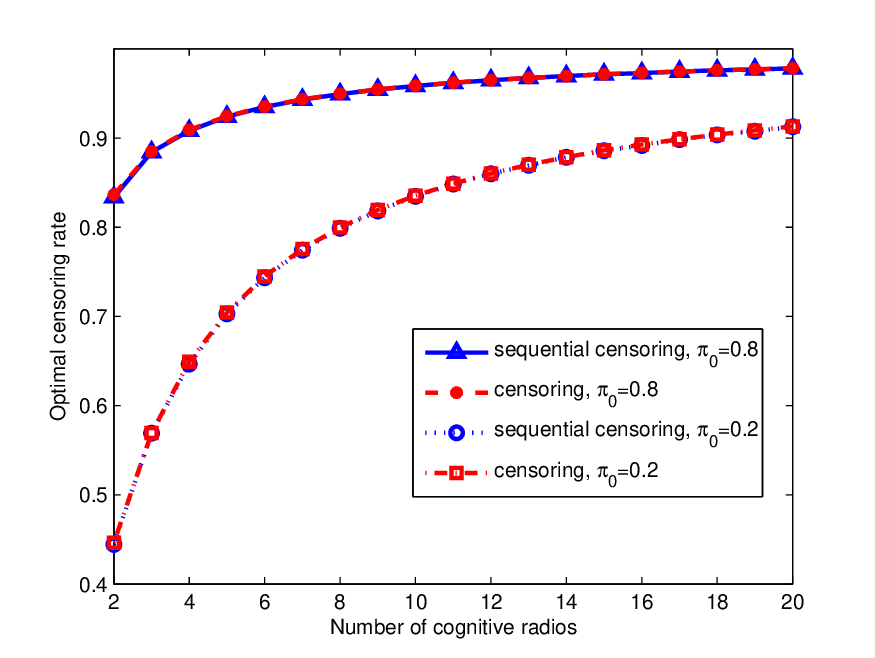}%
\label{rhom}}}
\hfil
\centerline{\subfloat[]{\includegraphics[width=3.2in, height=2.6in]{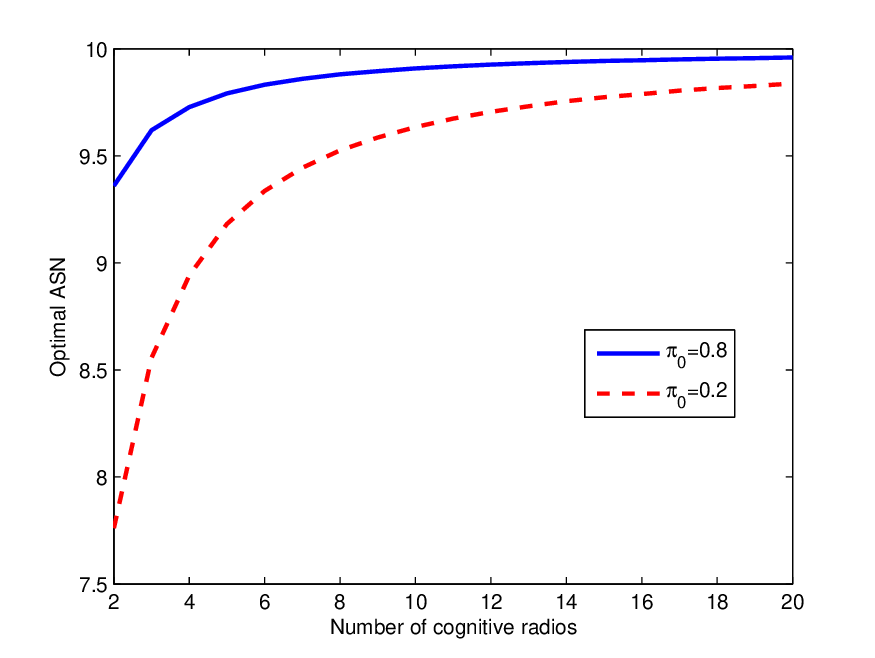}%
\label{asnm}}}
\caption{a)~Optimal maximum {average energy consumption per sensor} versus number of cognitive radios,~b)~Optimal censoring rate versus number of cognitive radios,~c)~Optimal ASN versus number of cognitive radios for the OR rule}
\end{figure}

Fig.~\ref{em} depicts the optimal maximum {average energy consumption per sensor} versus the number of cognitive radios for the OR rule. The SNR is assumed to be $0$~dB, $N=10$, $C_{s}=1$ and $C_{t}=10$. Furthermore, the probability of false alarm and detection constraints are assumed to be $\alpha=0.1$ and $\beta=0.9$ as determined by the IEEE 802.15.4 standard for cognitive radios \cite{SCSC}. It is shown for both high and low values of $\pi_{0}$ that censored sequential sensing outperforms the censoring scheme. Looking at Fig.~\ref{rhom} and Fig.~\ref{asnm}, where the respective optimal censoring rate and optimal ASN are shown versus the number of cognitive radios, we can deduce that the lower ASN is playing a key role in a lower energy consumption of the censored sequential sensing. Fig.~\ref{em} also shows that as the number of cooperating cognitive radios increases, the optimal maximum {average energy consumption per sensor} decreases and saturates, while as shown in Fig.~\ref{rhom} and Fig.~\ref{asnm}, the optimal censoring rate and optimal ASN increase. This way, the energy consumption tends to increase as a result of ASN growth and on the other hand inclines to decrease due to the censoring rate growth and that is the reason for saturation after a number of cognitive radios. Therefore, we can see that as the number of cognitive radios increases, a higher energy efficiency per sensor can be achieved. However, after a number of cognitive radios, the maximum {average energy consumption per sensor} remains almost at a constant level and by adding more cognitive radios no significant energy saving per sensor can be achieved while the total network energy consumption also increases.

\begin{figure}[!tp]
\centerline{\subfloat[]{\includegraphics[width=3.2in, height=2.6in]{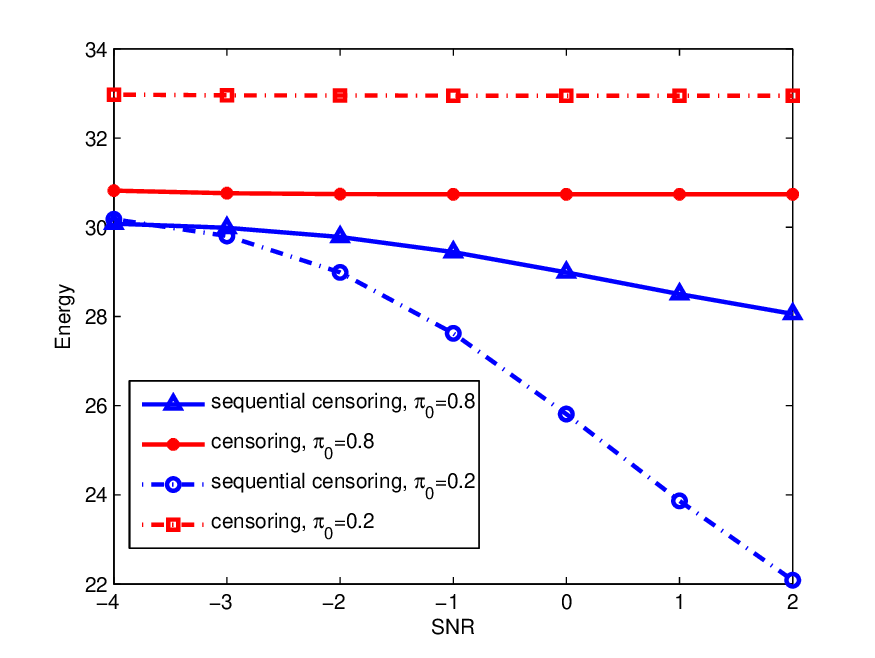}%
\label{esnr}}}
\hfil
\centerline{\subfloat[]{\includegraphics[width=3.2in, height=2.6in]{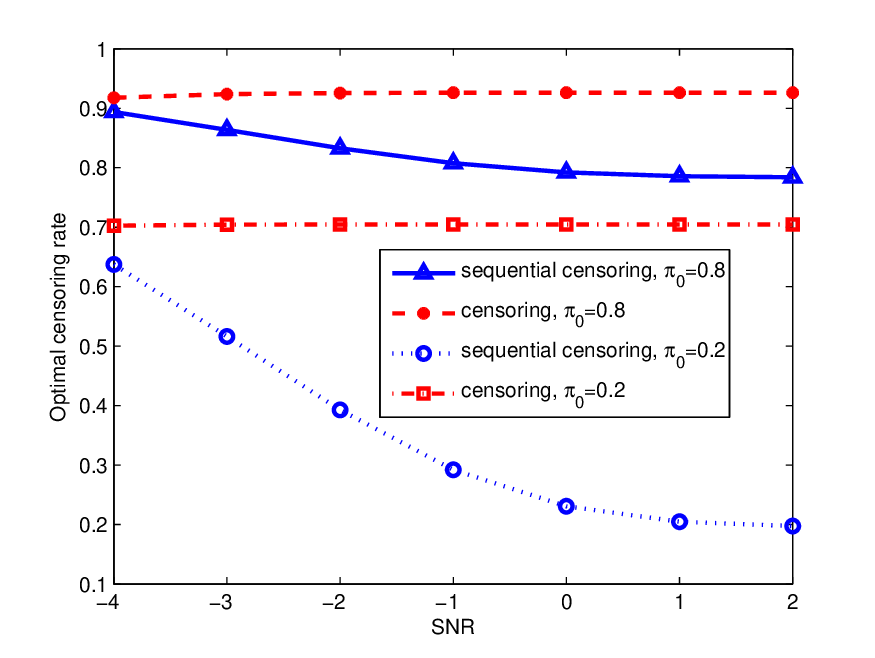}%
\label{rhosnr}}}
\hfil
\centerline{\subfloat[]{\includegraphics[width=3.2in, height=2.6in]{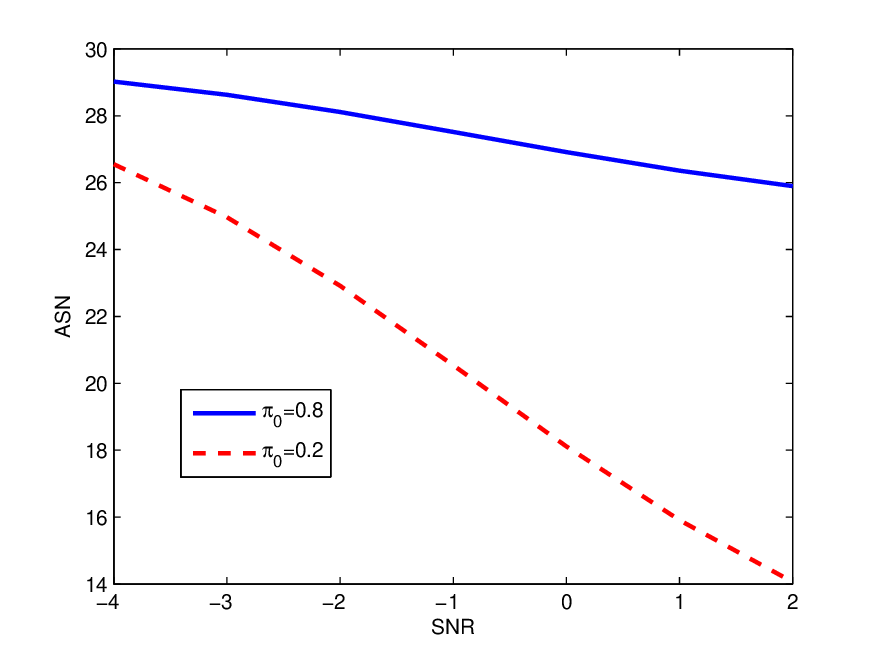}%
\label{asnsnr}}}
\caption{a)~Optimal maximum {average energy consumption per sensor} versus SNR,~b)~Optimal censoring rate versus SNR,~c)~Optimal ASN versus SNR for the OR rule}
\end{figure}

Figures~\ref{esnr},~\ref{rhosnr} and \ref{asnsnr} consider a scenario where $M=5$, $N=30$, $C_{sj}=1$, $C_{tj}=10$, $\alpha=0.1$, $\beta=0.9$ and $\pi_{0}$ can take a value of $0.2$ or $0.8$. The performance of the system versus SNR is analyzed in this scenario for the OR rule. The maximum {average energy consumption per sensor} is depicted in Fig.~\ref{esnr}. As for the earlier scenario, censored sequential sensing gives a higher energy efficiency compared to censoring. While the optimal energy variation for the censoring scheme is almost the same for all the considered SNRs, the censored sequential scheme's {average energy consumption per sensor} reduces significantly as the SNR increases. The reason is that as the SNR increases, the optimal ASN dramatically decreases (almost $50\%$ for $\gamma=2~$dB and $\pi_{0}=0.2$). This shows that as the SNR increases, censored sequential sensing becomes even more valuable and a significant energy saving per sensor can be achieved compared with the one that is achieved by censoring. {Since the SNR changes with the channel gain ($|h_{j}|^{2}$ under the first model or $\sigma_{hj}^{2}$ under the second model), from Fig.~\ref{esnr}, the behavior of the system with varying $|h_{j}|^{2}$ or $\sigma_{hj}^{2}$ can be derived, if the distribution of $|h_{j}|^{2}$ or $\sigma_{hj}^{2}$ is known.}

\begin{figure}[!tp]
\centerline{\subfloat[]{\includegraphics[width=3.2in, height=2.6in]{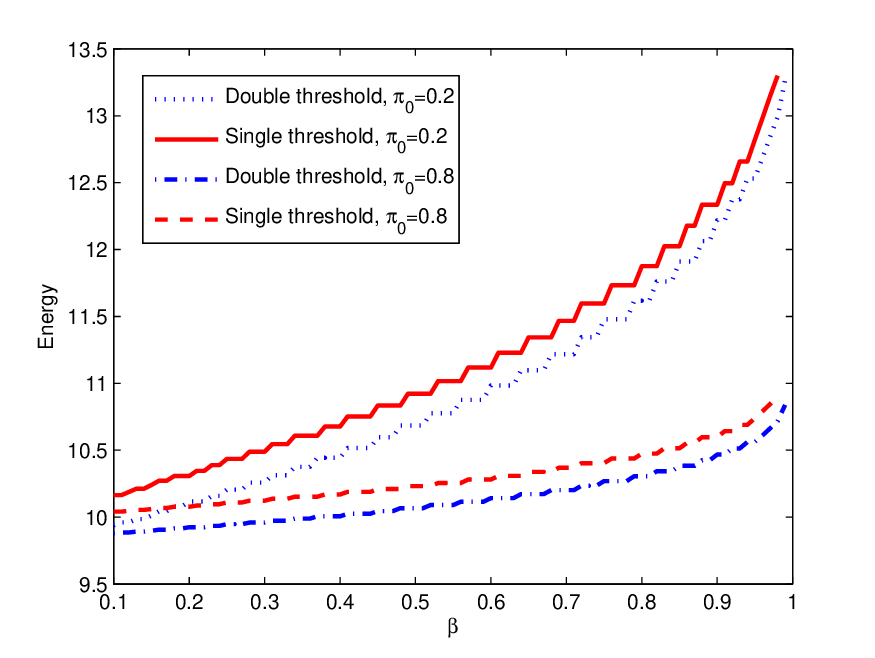}%
\label{epddbcs1}}}
\hfil
{\subfloat[]{\includegraphics[width=3.2in, height=2.6in]{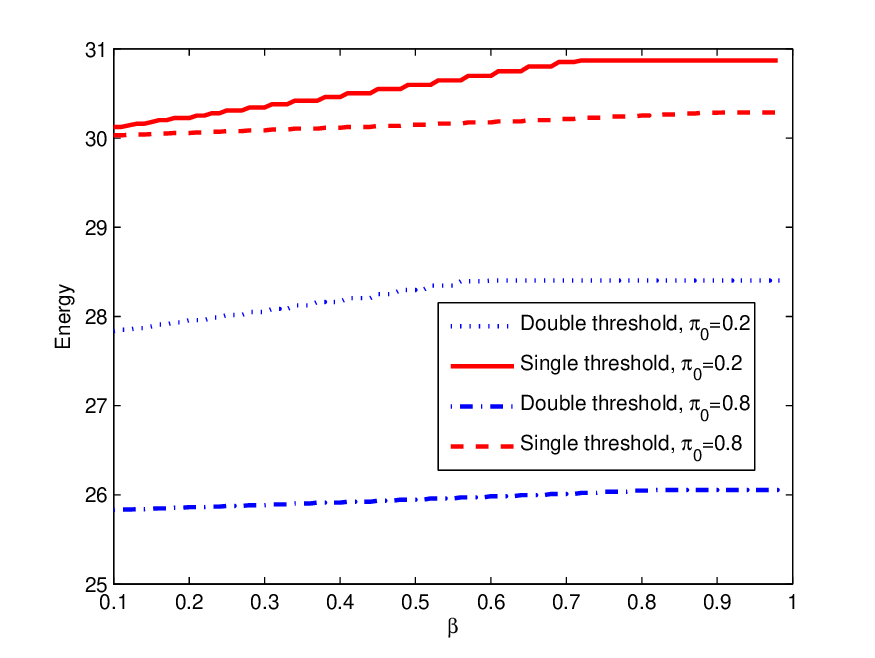}%
\label{epddbcs3}}}
\caption{Optimal maximum {average energy consumption per sensor} versus probability of detection constraint, $\beta$, for the OR rule,~a)~$C_s=1$,~b)~$C_s=3$}
\end{figure}

Figures~\ref{epddbcs1}~and~\ref{epddbcs3} compare the performance of the single threshold censored truncated sequential scheme with the one assuming two thresholds, i.e, $\bar{a}$ and $\bar{b}$ for the OR rule. The idea is to find when the double threshold scheme with its higher complexity becomes valuable. In these figures, $M=5$, $N=10$, $\gamma=0$~dB, $C_{t}=10$, $\pi_{0}=0.2,~0.8$, and $\alpha=0.1$, while $\beta$ changes from $0.1$ to $0.99$. The sensing energy per sample, $C_{s}$ in Fig.~\ref{epddbcs1} is assumed $1$, while in Fig.~\ref{epddbcs3} it is $3$. It is shown that as the sensing energy per sample increases, the energy efficiency of the double threshold scheme also increases compared to the one of the single threshold scheme, particularly when $\pi_{0}$ is high. The reason is that when $\pi_{0}$ is high, a much lower ASN can be achieved by the double threshold scheme compared to the single threshold one. This gain in performance comes at the cost of a {higher computational complexity because of the two-dimensional search}.

\begin{figure}[tp]
\centering {\includegraphics[width=3.2in, height=2.6in]{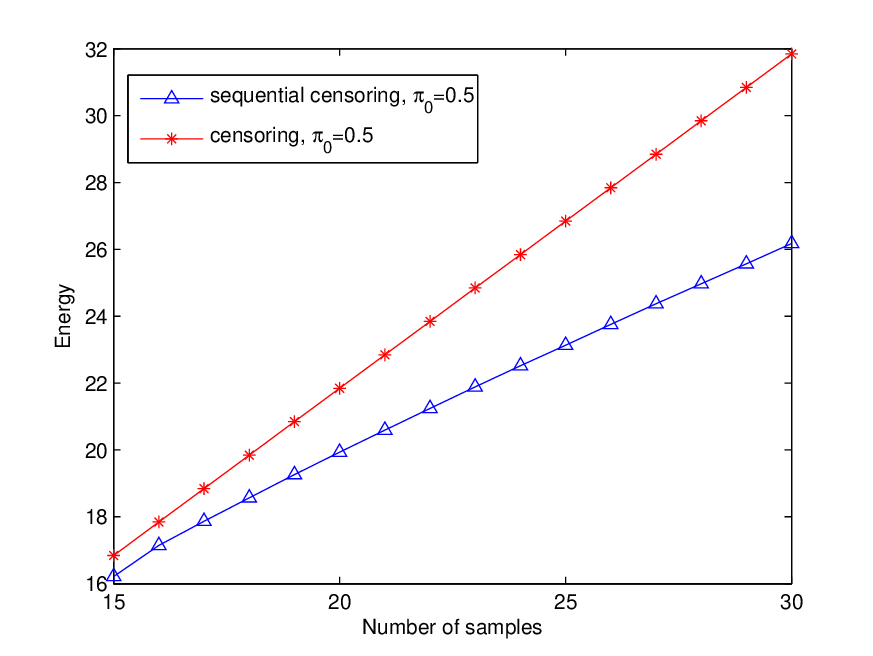}}
\caption{Optimal maximum {average energy consumption per sensor} versus number of samples for the OR rule} \label{en}
\end{figure}

{Fig.~\ref{en} depicts the optimal maximum {average energy consumption per sensor} versus the number of samples for the OR rule and for a network of $M=5$ cognitive radios where each radio experiences {a different channel gain and thus} a different SNR. Arranging the SNRs in a vector ${\bf{\gamma}}=[\gamma_{1},\dots,\gamma_{5}]$, we have ${\mathbf \gamma}=$[1dB, 2dB, 3dB, 4dB, 5dB]. The other parameters are $C_{s}=1$, $C_{t}=10$, $\pi_{0}=0.5$, $\alpha=0.1$ and $\beta=0.9$. As shown in Fig.~\ref{en}, by increasing the number of samples and thus the total sensing energy, the sequential censoring energy efficiency also increases compared to the censoring scheme. For example, if we define the efficiency of the censored truncated sequential sensing scheme as the difference of the optimal maximum average {energy consumption per sensor} of sequential censoring and censoring divided by the optimal maximum {average energy consumption per sensor} of censoring, the efficiency increases approximately three times from 0.06 (for $N=15$) to 0.19 (for $N=30$).}

\begin{figure}[tp]
\centering {\includegraphics[width=3.2in, height=2.6in]{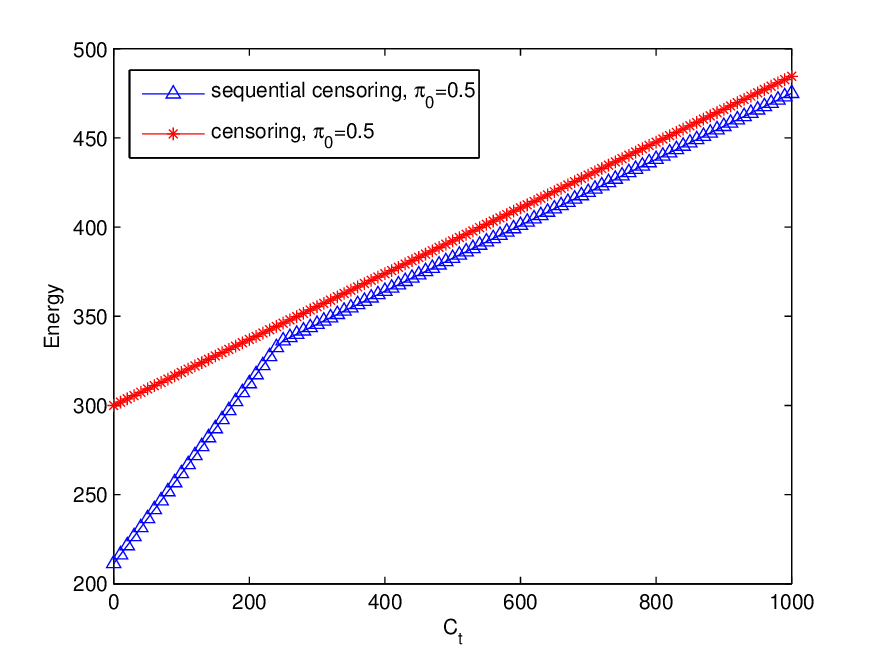}}
\caption{Optimal maximum {average energy consumption per sensor} versus transmission energy for the OR rule} \label{ect}
\end{figure}

{In Fig.~\ref{ect}, the sensing energy per sample is $C_{s}=10$ while the transmission energy $C_{t}$ changes from 0 to 1000. The goal is to see how the optimal maximum {average energy consumption per sensor} changes with $C_{t}$ for the or rule and for a network of $M=5$ cognitive radios with ${\boldsymbol{\gamma}}=$[1dB, 2dB, 3dB, 4dB, 5dB]. The other parameters of the network are $N=30$, $\pi_{0}=0.5$, $\alpha=0.1$ and $\beta=0.9$. The best saving for sequential censoring is achieved when the transmission energy is zero. Indeed, we can see that as the transmission energy increases the performance gain of sequential censoring reduces compared to censoring. However, in low-power radios where the sensing energy per sample and transmission energy are usually in the same range, sequential censoring performs much better than censoring in terms of energy efficiency as we can see in Fig.~\ref{ect}.} 

\begin{figure}[tp]
\centering {\includegraphics[width=3.2in, height=2.6in]{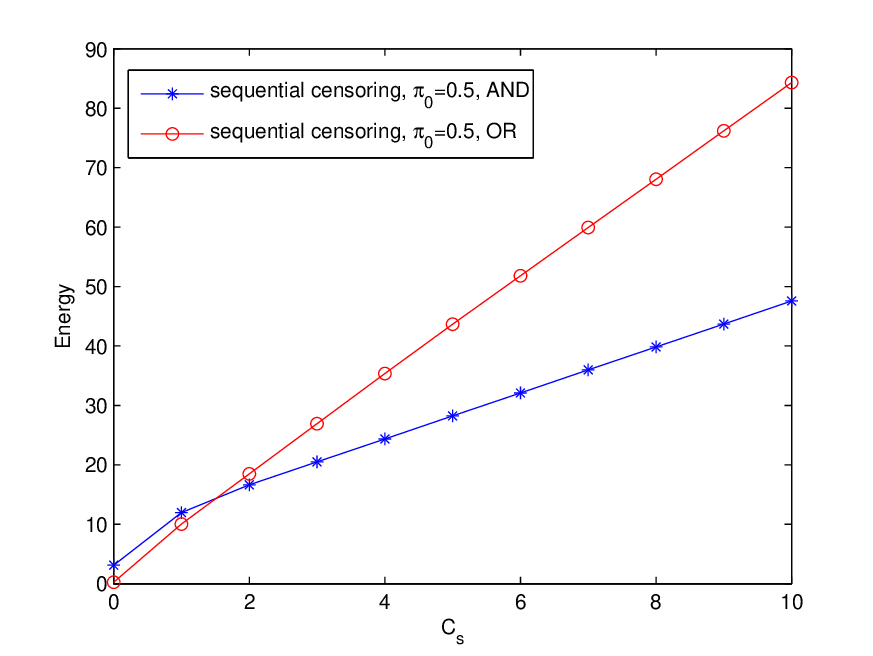}}
\caption{Optimal maximum {average energy consumption per sensor} versus sensing energy per sample for AND and OR rule} \label{ecs}
\end{figure}

{Fig.~\ref{ecs} depicts the optimal maximum average energy consumption per sensor versus the sensing energy per sample for both the AND and OR rule. For the sake of simplicity and tractability, the SNRs are assumed the same for $M=50$ cognitive radios. The other parameters are assumed to be $N=10$, $C_{t}=10$, $\pi_{0}=0.5$, $\gamma=0$~dB, $\alpha=0.1$ and $\beta=0.9$. For both fusion rules, the double threshold scheme is employed. We can see that the OR rule performs better for the low values of $C_{s}$. However, as $C_{s}$ increases the AND rule dominates and outperforms the OR rule, particularly for high values of $C_{s}$. The reason that the OR rule performs better than the AND rule at very low values of $C_{s}$ is that the optimal censoring rate for the OR rule is higher than the optimal censoring rate for the AND rule. However as $C_{s}$ increases, the AND rule dominates the OR rule in terms of energy efficiency due to the lower ASN.}

\begin{figure}[tp]
\centering {\includegraphics[width=3.2in, height=2.6in]{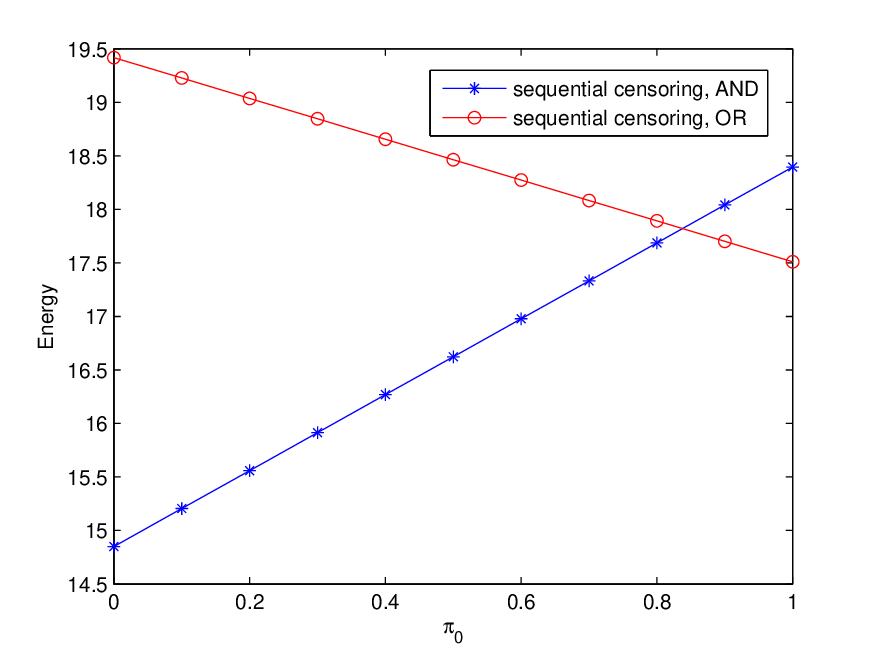}}
\caption{Optimal maximum {average energy consumption per sensor} versus $\pi_{0}$ for AND and OR rule} \label{eph0}
\end{figure}

{The optimal maximum average energy consumption per sensor versus $\pi_{0}$ is investigated in Fig.~\ref{eph0} for the AND and the OR rule. The underlying parameters are assumed to be $C_{s}=2$, $C_{t}=10$, $N=10$, $M=50$, $\gamma=0$~dB, $\alpha=0.1$ and $\beta=0.9$. It is shown that as the probability of the primary user absence increases, the optimal maximum average energy consumption per sensor reduces for the OR rule while it increases for the AND rule. This is mainly due to the fact that for the OR rule, we are mainly interested to receive a "1" from the cognitive radios. Therefore, as $\pi_{0}$ increases, the probability of receiving a "1" decreases, since the optimal censoring rate increases. The opposite happens for the AND rule, since for the AND rule, receiving a "0" from the cognitive radios is considered to be informative.}

\section{Summary and Conclusions}\label{conc}

We presented two energy efficient techniques for a cognitive sensor network. First, a censoring scheme has been discussed where each sensor employs a censoring policy to reduce the energy consumption. Then a censored truncated sequential approach has been proposed based on the combination of censoring and sequential sensing policies. We defined our problem as the minimization of the maximum {average energy consumption per sensor} subject to a global probability of false alarm and detection constraint {for the AND and the OR rules}. The optimal lower threshold is shown to be zero for the censoring scheme {in case of the OR rule while for the AND rule the optimal upper threshold is shown to be infinity}. Further, an explicit expression was given to find the optimal solution {for the OR rule and in case of the AND rule a closed for solution is derived}. We have further derived the analytical expressions for the underlying parameters in the censored sequential scheme and have shown that although the problem is not convex, a bounded two-dimensional search is possible {for both the OR rule and the AND rule}. Further, {in case of the OR rule,} we relaxed the lower threshold to obtain a line search problem in order to reduce the computational complexity. 

Different scenarios regarding transmission and sensing energy per sample as well as SNR, number of cognitive radios, {number of samples and detection performance constraints} were simulated for low and high values of $\pi_{0}$ { and for both the OR rule and the AND rule}. It has been shown that under the practical assumption of low-power radios, sequential censoring outperforms censoring. We conclude that for high values of the sensing energy per sample, despite its high computational complexity, the double threshold scheme developed for the OR rule becomes more attractive. { Further, it is shown that as the sensing energy per sample increases compared to the transmission energy, the AND rule performs better than the OR rule, while for very low values of the sensing energy per sample, the OR rule outperforms the AND rule.}

Note that a systematic solution for the censored sequential problem formulation was not given in this paper, and thus it is valuable to investigate a better algorithm to solve the problem. We also did not consider a combination of the proposed scheme with sleeping as in \cite{MPL3}, which can generate further energy savings. Our analysis was based on the OR rule {and the AND rule}, and thus extensions to other hard fusion rules could be interesting. 

\appendices

{\section{Optimal solution of (\ref{opt1})}\label{optpf}

Since the optimal $\lambda_{1}=0$, (\ref{delta0censexp}) and (\ref{delta1censexp}) can be simplified to $\delta_{0j}=1-P_{f}$ and $\delta_{1j}=1-P_{dj}$ and so (\ref{opt1}) becomes,
\begin{align} \label{opt2}
&\underset{\lambda_{2}}\min~\max_{j}\big[{~NC_{sj}+(\pi_{0}P_{f}+\pi_{1}P_{dj})C_{tj}}\big]\nonumber\\
 &\text{s.t.}~1-(1-P_{f})^{M}\le{\alpha},~1-\prod^{M}_{j=1}(1-P_{dj})\ge{\beta}.
  \end{align}

Since there is a one-to-one relationship between $\lambda_{2}$ and $P_{f}$, i.e., $\lambda_{2}=2\Gamma^{-1}[N,\Gamma(N)P_f]$ (where $\Gamma^{-1}$ is defined over the second argument), (\ref{opt2}) can be formulated as \cite[p.130]{BV},
\begin{equation} \label{opt3}
\begin{array}{lr}
\underset{P_{f}}\min~\max_{j}\big[{~NC_{sj}+(\pi_{0}P_{f}+\pi_{1}P_{dj})C_{tj}}\big]\\
 \text{s.t.}~1-(1-P_{f})^{M}\le{\alpha},~1-\prod^{M}_{j=1}(1-P_{dj})\ge{\beta}.
 \end{array}
 \end{equation}
Defining $P_{f}=F(\lambda_2)=\frac{\Gamma(N,\frac{\lambda_{2}}{2})}{\Gamma(N)}$ and $P_{dj}=G_{j}(\lambda_2)=\frac{\Gamma(N,\frac{\lambda_{2}}{2(1+\gamma_{j})})}{\Gamma(N)}$, we can write $P_{dj}$
as $P_{dj}=G_{j}(F^{-1}(P_{f}))$. Calculating the derivative of
$C_{j}$ with respect to $P_{f}$, we find that
\begin{equation}
\frac{\partial{C_{j}}}{\partial{P_{f}}}=\frac{\partial{\big[C_{tj}(\pi_0 P_{f}+\pi_1 P_{dj})\big]}}{\partial{P_{f}}}=C_{tj}\pi_0+~\frac{\partial{P_{dj}}}{\partial{P_f}}\ge{0},
\end{equation}
where we use the fact that
\begin{align} \frac{\partial{P_{dj}}}{\partial{P_f}}&=~\frac{-\frac{1}{2^{N}\Gamma(N)}{2\Gamma^{-1}[N,\Gamma(N)P_f]^{N-1}e^{2\Gamma^{-1}[N,\Gamma(N)P_f]/2(1+\gamma_{j})}I_{\{2\Gamma^{-1}[N,\Gamma(N)P_f]\ge{0}\}}}}{-\frac{1}{2^{N}\Gamma(N)}{2\Gamma^{-1}[N,\Gamma(N)P_f]^{N-1}e^{2\Gamma^{-1}[N,\Gamma(N)P_f]/2}I_{\{2\Gamma^{-1}[N,\Gamma(N)P_f]\ge{0}\}}}}\nonumber\\
&=e^{2\Gamma^{-1}[N,\Gamma(N)P_f](1/2(1+\gamma_{j})-1/2)}\ge{0}.
\end{align}

Therefore, we can simplify
(\ref{opt3}) as 
\begin{equation} \label{opt4}
\begin{array}{lr}
\underset{P_{f}}\min~P_{f}\\
 \text{s.t.}~1-(1-P_{f})^{M}\le{\alpha},~1-\prod^{M}_{j=1}(1-P_{dj})\ge{\beta}.
 \end{array}
 \end{equation}
which can be easily solved by a line search over $P_{f}$. However, since $Q^{c}_{\text{D}}$ is a monotonically increasing function of $P_{f}$, i.e., $Q^{c}_{\text{D}}=H(P_{f})=1-\prod^{M}_{j=1}(1-G_{j}(F^{-1}(P_{f})))$ and thus $\frac{\partial{Q^{c}_{\text{D}}}}{\partial{P_{f}}}=~\frac{\partial{Q^{c}_{\text{D}}}}{\partial{P_{dj}}}\frac{\partial{P_{dj}}}{\partial{P_{f}}}=\prod^{l=M}_{l=1,l\ne{j}}(1-P_{dl})\frac{\partial{P_{dj}}}{\partial{P_{f}}}\ge{0}$, we can further simplify the constraints in (\ref{opt4}) as $P_{f}\le{1-(1-\alpha)^{1/M}}$ and $P_{f}\ge{H^{-1}(\beta)}$. Thus, we obtain
 \begin{equation} \label{opt5}
 \begin{array}{lr}
\underset{P_{f}}\min~P_{f}\\
 \text{s.t.}~P_{f}\le{1-(1-\alpha)^{1/M}},~P_{f}\ge{H^{-1}(\beta)}.
 \end{array}
 \end{equation}
Therefore, if the feasible set of (\ref{opt5}) is not empty, then the optimal solution is given by $P_{f}=H^{-1}(\beta)$.}

\section{Derivation of $Pr(E_{n}|{\cal H}_{0})$}\label{prenh0}

Introducing $\Gamma_{n}=\{a_{i}<\zeta_{ij}<b_{i},~i=1,...,n-1\}$ and $p_{n}=\frac{1}{2^{n-1}}e^{-b_{n}/2}$, we can write
\begin{eqnarray}
Pr(E_{n}|{\cal H}_{0})&=&\underset{\Gamma_{n}}{\int...\int}\int^{\infty}_{b_{n}}\frac{1}{2^{n}}e^{-\zeta_{nj}/2}I_{\{0\le{{\zeta_{1j}}\le{\zeta_{2j}}...\le{\zeta_{nj}}\}}}d\zeta_{1j}...d\zeta_{nj}\nonumber\\
&=&p_{n}\underset{\Gamma_{n}}{\int...\int}I_{\{0\le{\zeta_{1j}}\le{\zeta_{2j}}...\le{\zeta_{n-1j}}\}}d\zeta_{1j}...d\zeta_{n-1j}.
\end{eqnarray}
Denoting $A(n)=\underset{\Gamma_{n}}{\int...\int}I_{\{0\le{{\zeta_{1j}}\le{\zeta_{2j}}...\le{\zeta_{n-1j}}\}}d\zeta_{1j}...d\zeta_{n-1j}}$, we obtain
\begin{align}\label{an}
A(n)=\left\{
\begin{array}{lr}
\frac{b_{1}b^{n-2}_{n}}{(n-1)!},~n=1,...,p+1\\
\big[f^{(n-1)}_{\boldsymbol{a}^{n-1}_{0}}(b_{n-1})-I_{\{n\ge{3}\}}\sum^{n-3}_{i=0}\frac{(b_{n-1}-b_{i+1})^{n-i-1}}{(n-i-1)!}2^{i}e^{\frac{b_{i+1}}{2}}Pr(E_{i+1}|{\cal H}_{0})\big],~n=p+2,...,q+1\\
\big[f^{(n-1)}_{\boldsymbol{a}^{n-1}_{0}}(b_{n-1})-\sum^{n}_{i=0}f^{(n-1-i)}_{\boldsymbol{\psi}^{n-1}_{i,a_{n-1}}}(b_{n-1})2^{i}e^{\frac{b_{i+1}}{2}}Pr(E_{i+1}|{\cal H}_{0})\big],~n=q+2,...,N
\end{array}, \right.
\end{align}
where $\boldsymbol{a}^{n-1}_{0}=[a_{0},\dots,a_{n-1}]$. Denoting $q$ to be the smallest integer for which $a_{q}\le{b_{1}}<b_{q}$, and $c$ and $d$ to be two non-negative real numbers satisfying $0\le{c}<d$, $a_{n-1}\le{c}\le{b_{n}}$ and $a_{n}\le{d}$, $\eta_{0}=0,~\boldsymbol{\eta}_{k}=[\eta_{1},...,\eta_{k}],~0\le{\eta_{1}}\le...\le{\eta_{k}}$, the functions $f^{(k)}_{\boldsymbol{\eta}_{k}}(\zeta)$ and the vector $\boldsymbol{\psi}^{n}_{i,c}$ in (\ref{an}) are as follows
\begin{eqnarray}\label{f}
&f^{(k)}_{\boldsymbol{\eta}_{k}}(\zeta)=\sum^{k-1}_{i=0}\frac{f^{(k)}_{i}(\zeta-\eta_{i+1})^{k-i}}{(k-i)!}+f^{(k)}_{k}\nonumber\\
&f^{(k)}_{i}=f^{(k-1)}_{i},~i=0,...,k-1,~k\ge{1},~f^{(k)}_{k}=-\sum^{k-1}_{i=0}\frac{f^{(k-1)}_{i}}{(k-i)!}(\eta_{k}-\eta_{i+1})^{k-i},~f^{(0)}_{0}=1,
\end{eqnarray}
\begin{eqnarray}\label{psi}
\boldsymbol{\psi}^{n}_{i,c}=\left\{
\begin{array}{lr}
[\underset{q}{\underbrace{b_{i+1},...,b_{i+1}}},\underset{n-q-i}{\underbrace{a_{q+i+1},...,a_{n-1},c}}],~i\in[0,n-q-2]\\
{\underset{n-i}{\underbrace{[b_{i+1},...,b_{i+1},c}}]},~i\in[n-q-1,s-1]\\
b_{i+1}\boldsymbol{1}_{n-i},~i\in[s,n-2]\end{array}, \right.
\end{eqnarray}
with $s$ denoting the integer for which $b_{s}<c\le{b_{s+1}}$ and $f^{(0)}_{\boldsymbol{\eta}_{k}}(\zeta)=1$.

\section{Derivation of $Pr(E_{n}|{\cal H}_{1})$}\label{prenh1}

Introducing $q_{n}=\frac{1}{[2(1+\gamma_{j})]^{n-1}}e^{-b_{n}/2(1+\gamma_{j})}$, we can write
\begin{eqnarray}
Pr(E_{n}|{\cal H}_{1})&=&\underset{\Gamma_{n}}{\int...\int}\int^{\infty}_{b_{n}}\frac{1}{[2(1+\gamma_{j})]^{n}}e^{-\zeta_{nj}/2(1+\gamma_{j})}I_{\{0\le{{\zeta_{1j}}\le{\zeta_{2j}}...\le{\zeta_{nj}}\}}}d\zeta_{1j}...d\zeta_{nj}\nonumber\\
&=&q_{n}\underset{\Gamma_{n}}{\int...\int}I_{\{0\le{{\zeta_{1j}}\le{\zeta_{2j}}...\le{\zeta_{n-1j}}\}}}d\zeta_{1j}...d\zeta_{n-1j}.
\end{eqnarray}
Denoting $B(n)=\underset{\Gamma_{n}}{\int...\int}I_{\{0\le{{\zeta_{1j}}\le{\zeta_{2j}}...\le{\zeta_{n-1j}}\}}d\zeta_{1j}...d\zeta_{n-1j}}$, and using the notations of Appendix~\ref{prenh0}, we obtain
\begin{align}
\nonumber\\
B(n)=\left\{
\begin{array}{lr}
\frac{b_{1}b^{n-2}_{n}}{(n-1)!},~n=1,...,p+1\\
\big[f^{(n-1)}_{\boldsymbol{a}^{n-1}_{0}}(b_{n-1})-I_{\{n\ge{3}\}}\sum^{n-3}_{i=0}\frac{(b_{n-1}-b_{i+1})^{n-i-1}}{(n-i-1)!}[2(1+\gamma_{j})]^{i}e^{\frac{b_{i+1}}{2(1+\gamma_{j})}}Pr(E_{i+1}|{\cal H}_{1})\big],~n=p+2,...,q+1\\
\big[f^{(n-1)}_{\boldsymbol{a}^{n-1}_{0}}(b_{n-1})-\sum^{n-3}_{i=0}f^{(n-1-i)}_{\boldsymbol{\psi}^{n-1}_{i,a_{n-1}}}(b_{n-1})[2(1+\gamma_{j})]^{i}e^{\frac{b_{i+1}}{2(1+\gamma_{j})}}Pr(E_{i+1}|{\cal H}_{1})\big],~n=q+2,...,N
\end{array}. \right.
\end{align}

\section{Analytical expression for $J^{(n)}_{a_{n},b_{n}}(\theta)$}\label{j}

Under $\theta>0,~n\ge{1}$ and $0\le{\zeta_{1j}}\le...\le{\zeta_{nj}},~\zeta_{ij}\in(a_{i},b_{i}),~i=1,...,n$, the function $J^{(n)}_{a_{n},b_{n}}(\theta)$ is defined as \cite{XZ}
\begin{equation}
J^{(n)}_{a_{n},b_{n}}(\theta)=\sum^{n}_{i=1}\theta^{-i}\big[f^{(n-i)}_{\boldsymbol{a}^{n-i}_{0}}(a_{n})e^{-\theta a_{n}}-f^{(n-i)}_{\boldsymbol{a}^{n-i}_{0}}(b_{n})e^{-\theta b_{n}}\big]-I_{\{n\ge{2}\}}\sum^{n-2}_{k=0}g^{(k)}_{a_{n},b_{n}}(\theta),
\end{equation}
where using the notations of Appendix~\ref{prenh0}, we have \cite{XZ}
\begin{equation}
g^{(k)}_{c,d}=\left\{\
\begin{array}{lr}
I^{(k)}\big[\theta^{k-n}e^{-\theta b_{k+1}}-\sum^{n-k}_{i=1}\theta^{-i}f^{(n-k-i)}_{b_{k+1}\boldsymbol{1}_{n-k-i}}(d)e^{-\theta d}\big],~c\le{b_{1}},~k\in{[0,n-2]}\\
I^{(k)}\sum^{n-k}_{i=1}\theta^{-i}\big[f^{(n-k-i)}_{\boldsymbol{\psi}^{n-i}_{k,c}}(c)e^{-\theta c}-f^{(n-k-i)}_{\boldsymbol{\psi}^{n-i}_{k,d}}(d)e^{-\theta d}\big],~c>b_{1},~k\in{[0,s-1]}\\
I^{(k)}\big[\theta^{k-n}e^{-\theta b_{k+1}}-\sum^{n-k}_{i=1}\theta^{-i}f^{(n-k-i)}_{b_{k+1}\boldsymbol{1}_{n-k-i}}(d)e^{-\theta d}\big],~c>b_{1},~k\in{[s,n-2]}
\end{array},\right.
\end{equation}
with $I^{(0)}=1$ and
\begin{equation}
I^{(n)}=\left\{
\begin{array}{lr}
f^{(n)}_{\boldsymbol{a}^{n}_{0}}(b_{n})-I_{\{n\ge{2}\}}\sum^{n-2}_{i=0}\frac{(b_{n}-b_{i+1})^{n-i}}{(n-i)!}I^{(i)},~n\in{[1,q]}\\
f^{(n)}_{\boldsymbol{a}^{n}_{0}}(b_{n})-\sum^{n-2}_{i=0}f^{(n-i)}_{\boldsymbol{\psi}^{n}_{i,a_{n}}}(b_{n})I^{(i)},~n\in{[q+1,\infty)}
\end{array}.\right.
\end{equation}

\section{Proof of Theorem 1}\label{thr1}
Assume that $P_{f}$ and $P_{d}$ are the respective given local probability of false alarm and detection. Denoting $\rho^{c}$ as the censoring rate for the optimal censoring scheme (\ref{opt5}), we obtain $1-\rho^{c}=\pi_{0}P_{f}+\pi_{1}P_{d}$, and denoting $\rho^{cs}$ as the censoring rate for the optimal censored truncated sequential sensing (\ref{probform}), based on what we have discussed in Section~\ref{censprob}, we obtain $1-\rho^{cs}=\pi_{0}(P_{f}+\mathcal{L}_{0}(\bar{a},\bar{b}))+\pi_{1}(P_{d}+\mathcal{L}_{1}(\bar{a},\bar{b}))$. Note that $\mathcal{L}_{k}(\bar{a},\bar{b}),~k=0,1$, represents the probability that $\zeta_{n}\le{a_{n}},~n=1,\dots,N$ under ${\cal H}_{k}$ which is non-negative. Hence, we can conclude that $1-\rho^{cs}\ge 1-\rho^{c}$ and thus $\rho^{c}\ge{\rho^{cs}}$.

{\section{Optimal solution of (\ref{andopt})}\label{optdelta0}

Since the optimal $\lambda_{2}\rightarrow{\infty}$, (\ref{qfand}) and (\ref{qdand}) can be simplified to $Q^{c}_{\text{F,AND}}=\delta^{M}_{0}$ and $Q^{c}_{\text{D,AND}}=\prod^{M}_{j=1}\delta_{1j}$ and so (\ref{andopt}) becomes,
\begin{align} \label{optand2}
&\underset{\lambda_{1}}\min~\max_{j}\big[{~NC_{sj}+(\pi_{0}(1-\delta_{0})+\pi_{1}(1-\delta_{1j}))C_{tj}}\big]\nonumber\\
 &\text{s.t.}~\delta_{0}^{M}\le{\alpha},~\prod^{M}_{j=1}\delta_{1j}\ge{\beta}.
  \end{align}

Since there is a one-to-one relationship between $\lambda_{1}$ and $\delta_{0}$, i.e., $\lambda_{1}=2\Gamma^{-1}[N,\Gamma(N)\delta_{0}]$ (where $\Gamma^{-1}$ is defined over the second argument), (\ref{optand2}) can be formulated as \cite[p.130]{BV},
\begin{equation} \label{optand3}
\begin{array}{lr}
\underset{\delta_{0}}\min~\max_{j}\big[{~NC_{sj}+(\pi_{0}(1-\delta_{0})+\pi_{1}(1-\delta_{1j}))C_{tj}}\big]\\
 \text{s.t.}~\delta^{M}_{0}\le{\alpha},~\prod^{M}_{j=1}\delta_{1j}\ge{\beta}.
 \end{array}
 \end{equation}
Defining $\delta_{0}=F_{\text{AND}}(\lambda_1)=\frac{\Gamma(N,\frac{\lambda_{1}}{2})}{\Gamma(N)}$ and $\delta_{1j}=G_{\text{AND,j}}(\lambda_1)=\frac{\Gamma(N,\frac{\lambda_{1}}{2(1+\gamma_{j})})}{\Gamma(N)}$, we can write $\delta_{1j}$
as $\delta_{1j}=G_{\text{AND},j}(F^{-1}(\delta_{0}))$. Calculating the derivative of
$C_{j}$ with respect to $\delta_{0}$, we find that
\begin{equation}
\frac{\partial{C_{j}}}{\partial{\delta_{0}}}=\frac{\partial{\big[C_{tj}(\pi_0 (1-\delta_{0})+\pi_1 (1-\delta_{1j}))\big]}}{\partial{\delta_{0}}}=-C_{tj}\pi_0+~\frac{\partial{(1-\delta_{1j})}}{\partial{\delta_{0}}}\le{0},
\end{equation}
where we use the fact that
\begin{align} \frac{\partial{\delta_{1j}}}{\partial{\delta_{0}}}&=~\frac{-\frac{1}{2^{N}\Gamma(N)}{2\Gamma^{-1}[N,\Gamma(N)\delta_{0}]^{N-1}e^{2\Gamma^{-1}[N,\Gamma(N)\delta_{0}]/2(1+\gamma_{j})}I_{\{2\Gamma^{-1}[N,\Gamma(N)\delta_{0}]\ge{0}\}}}}{-\frac{1}{2^{N}\Gamma(N)}{2\Gamma^{-1}[N,\Gamma(N)\delta_{0}]^{N-1}e^{2\Gamma^{-1}[N,\Gamma(N)\delta_{0}]/2}I_{\{2\Gamma^{-1}[N,\Gamma(N)\delta_{0}]\ge{0}\}}}}\nonumber\\
&=e^{2\Gamma^{-1}[N,\Gamma(N)\delta_{0}](1/2(1+\gamma_{j})-1/2)}\ge{0}.
\end{align}

Therefore, we can simplify
(\ref{optand3}) as 
\begin{equation} \label{optand4}
\begin{array}{lr}
\underset{\delta_{0}}\max~\delta_{0}\\
 \text{s.t.}~\delta^{M}_{0}\le{\alpha},~\prod^{M}_{j=1}\delta_{1j}\ge{\beta}.
 \end{array}
 \end{equation}
Since $Q^{c}_{\text{D,AND}}$ is a monotonically increasing function of $\delta_{0}$, i.e., $Q^{c}_{\text{D,AND}}=H_{\text{AND}}(\delta_{0})=\prod^{M}_{j=1}(G_{\text{AND},j}(F_{\text{AND}}^{-1}(\delta_{0})))$ and thus $\frac{\partial{Q^{c}_{\text{D,AND}}}}{\partial{\delta_{0}}}=~\frac{\partial{Q^{c}_{\text{D,AND}}}}{\partial{\delta_{1j}}}\frac{\partial{\delta_{1j}}}{\partial{\delta_{0}}}=\prod^{l=M}_{l=1,l\ne{j}}(\delta_{1l})\frac{\partial{\delta_{1j}}}{\partial{\delta_{0}}}\ge{0}$, we can further simplify the constraints in (\ref{optand4}) as $\delta_{0}\le{\alpha^{1/M}}$ and $\delta_{1j}\ge{H^{-1}(\beta)}$. Thus, we obtain
 \begin{equation} \label{optand5}
\begin{array}{lr}
\underset{\delta_{0}}\max~\delta_{0}\\
 \text{s.t.}~\delta_{0}\le{\alpha^{1/M}},~\delta_{1j}\ge{H^{-1}(\beta)}.
 \end{array}
 \end{equation}
Therefore, if the feasible set of (\ref{optand5}) is not empty, then the optimal solution is given by $\delta_{0}=\alpha^{1/M}(\beta)$.}

\begin{IEEEbiography}[{\includegraphics[width=1in,height=1.25in,clip,keepaspectratio]{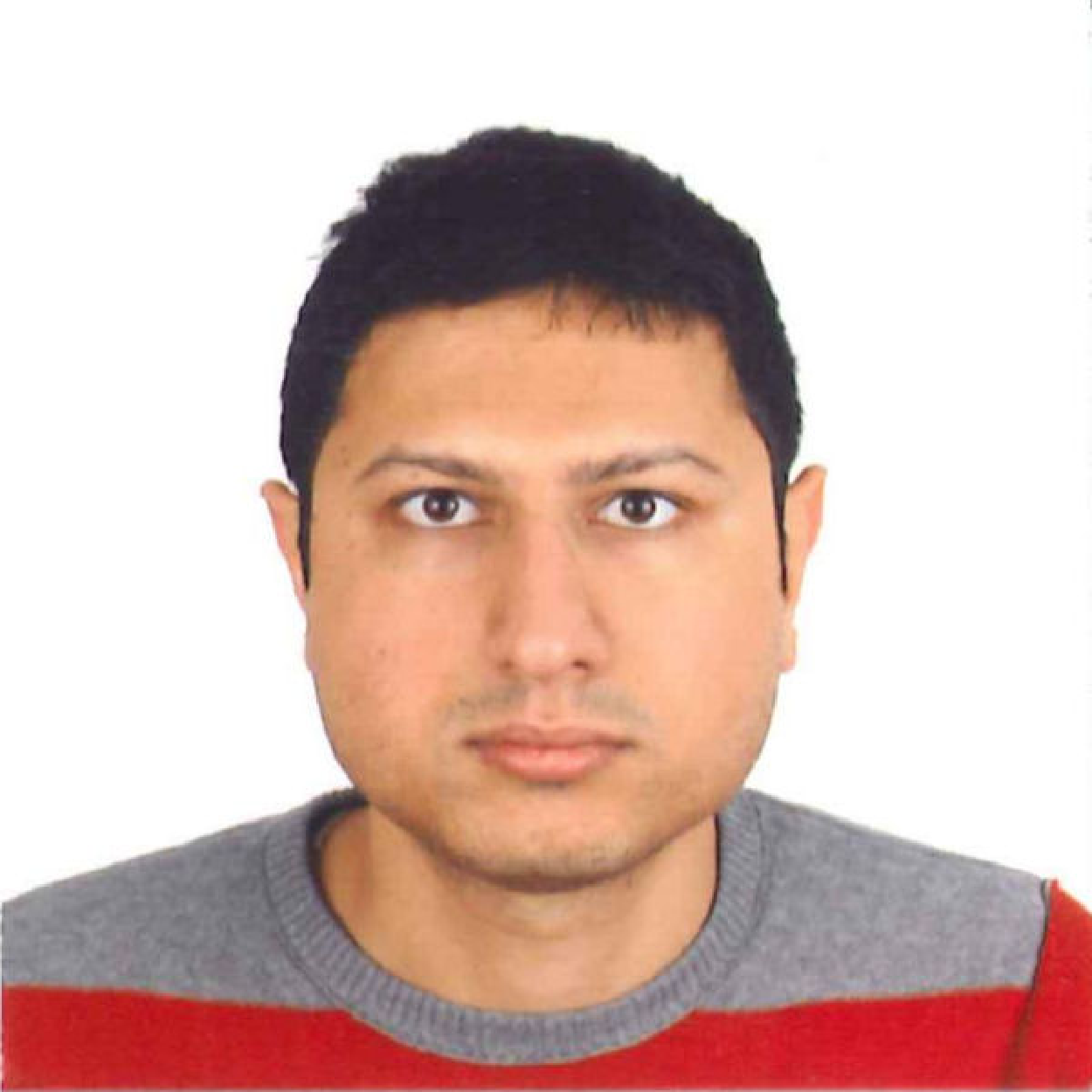}}]{Sina Maleki}
received his B.Sc. degree in electrical
engineering from Iran University of Science
and Technology, Tehran, Iran, in 2006, and his
M.S. degree in electrical engineering from Delft University of Technology,
Delft, The Netherlands, in 2009. From July
2008 to April 2009, he was an intern student
at the Philips Research Center, Eindhoven, The
Netherlands, working on spectrum sensing for
cognitive radio networks. He then joined the Circuits
and Systems Group at the Delft University
of Technology, where he is currently a Ph.D. student. He has served as a
reviewer for several journals and conferences.
\end{IEEEbiography}
 
\begin{IEEEbiography}[{\includegraphics[width=1in,height=1.25in,clip,keepaspectratio]{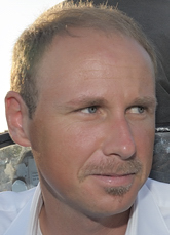}}]{Geert Leus}
was born in Leuven, Belgium, in 1973. He received the
electrical engineering degree and the PhD degree in applied sciences 
from the Katholieke Universiteit Leuven, Belgium, in June 1996 and
May 2000, respectively. He has been a Research Assistant and a
Postdoctoral Fellow of the Fund for Scientific Research - Flanders,
Belgium, from October 1996 till September 2003. During that period,
Geert Leus was affiliated with the Electrical Engineering Department
of the Katholieke Universiteit Leuven, Belgium. Currently, Geert Leus
is an Associate Professor at the Faculty of Electrical Engineering,
Mathematics and Computer Science of the Delft University of
Technology, The Netherlands. His research interests are in the area of 
signal processing for communications. Geert Leus received a 2002 IEEE 
Signal Processing Society Young Author Best Paper Award and a 2005 IEEE 
Signal Processing Society Best Paper Award. He was the Chair of the IEEE 
Signal Processing for Communications and Networking Technical Committee, 
and an Associate Editor for the IEEE Transactions on Signal Processing, 
the IEEE Transactions on Wireless Communications, and the IEEE Signal 
Processing Letters. Currently, he is a member of the IEEE Sensor Array 
and Multichannel Technical Committee and serves on the Editorial Board 
of the EURASIP Journal on Advances in Signal Processing. Geert Leus has 
been elevated to IEEE Fellow.
\end{IEEEbiography}

\end{document}